\documentclass[fleqn,usenatbib]{mnras}

\usepackage{newtxtext,newtxmath}
\usepackage[T1]{fontenc}

\DeclareRobustCommand{\VAN}[3]{#2}
\let\VANthebibliography\thebibliography
\def\thebibliography{\DeclareRobustCommand{\VAN}[3]{##3}\VANthebibliography}

\usepackage{graphicx}	% Including figure files
\usepackage{amsmath}	% Advanced maths commands
\usepackage{multirow} % \multirow{10}{*}{3}
\usepackage{threeparttable}

\newcommand{\ha}{\hbox{H$\alpha$}}
\newcommand{\hb}{\hbox{H$\beta$}}

\newcommand \Lsun{L_\odot}
\newcommand \kms{\, {\rm km \, s}^{-1} }

\def\gtsim{~\rlap{$>$}{\lower 1.0ex\hbox{$\sim$}}}
\def\ltsim{~\rlap{$<$}{\lower 1.0ex\hbox{$\sim$}}}

\title[Star formation in NGC 1808]{Star formation in the centre of NGC 1808 as observed by ALMA}

\author[Chen et al.]{
Guangwen Chen$^{1,2,3}$\thanks{E-mail: guangwen@mail.ustc.edu.cn}, % guangwen.chen@manchester.ac.uk
George J. Bendo$^{4}$, %\thanks{E-mail: george.bendo@manchester.ac.uk}
Gary A. Fuller$^{4,5}$,
Christian Henkel$^{6,7}$,
and 
Xu Kong$^{1,2,8}$
\\
$^{1}$Deep Space Exploration Laboratory / Department of Astronomy, University of Science and Technology of China, Hefei 230026, China\\
$^{2}$School of Astronomy and Space Sciences, University of Science and Technology of China, Hefei 230026, China\\
$^{3}$Jodrell Bank Centre for Astrophysics, Department of Physics and Astronomy, The University of Manchester, Oxford Road, Manchester M13 9PL, United Kingdom\\
$^{4}$UK ALMA Regional Centre Node, Jodrell Bank Centre for Astrophysics, Department of Physics and Astronomy, The University of Manchester,\\ Oxford Road, Manchester M13 9PL, United Kingdom\\
$^{5}$I. Physikalisches Institut, University of Cologne, Z\"ulpicher Str. 77, 50937 K\"oln, Germany\\
$^{6}$Max-Planck-Institut f\"ur Radioastronomie, Auf dem H\"ugel 69, D-53121 Bonn, Germany \\
$^{7}$Astronomy Department, Faculty of Science, King Abdulaziz University, P.O. Box 80203, Jeddah 21589, Saudi Arabia \\
$^{8}$Frontiers Science Center for Planetary Exploration and Emerging Technologies, University of Science and Technology of China, Hefei, Anhui, 230026, China \\
}

\date{Accepted XXX. Received YYY; in original form ZZZ}

\pubyear{2023}

\begin{document}
\label{firstpage}
\pagerange{\pageref{firstpage}--\pageref{lastpage}}
\maketitle

\begin{abstract}

We present Atacama Large Millimeter/submillimeter Array (ALMA) observations of 85.69 and 99.02 GHz continuum emission and H42$\alpha$ and H40$\alpha$ lines emission from the central 1~kpc of NGC 1808.  These forms of emission are tracers of photoionizing stars but unaffected by dust obscuration that we use to test the applicability of other commonly star formation metrics.   An analysis of the spectral energy distributions shows that free-free emission contributes about 60 to 90 per cent of the continuum emission in the 85-100 GHz frequency range, dependent on the region.  
%The average electron temperature for the central starburst is of $\sim$5000~K, which is comparable to that of the centre of the Milky Way.  
The star formation rate (SFR) derived from the ALMA free-free emission is $3.1\pm0.3$~M$_\odot$~yr$^{-1}$.  This is comparable to the SFRs measured from the infrared emission, mainly because most of the bolometric energy from the heavily obscured region is emitted as infrared emission.  The radio 1.5~GHz emission yields a SFR 25 per cent lower than the ALMA value, probably because of the diffusion of the electrons producing the synchrotron emission beyond the star-forming regions.  The SFRs measured from the extinction-corrected $\ha$ line emission are about 40 to 65 per cent of the SFR derived from the ALMA data, likely because this metric was not calibrated for high extinction regions.  Some SFRs based on extinction-corrected ultraviolet emission are similar to those from ALMA and infrared data, but given that the ultraviolet terms in the extinction correction equations are very small, these metrics seem inappropriate to apply to this dusty starburst.  
% In addition, our results suggest that any AGN that may be present in NGC~1808 can only produce no more than 20 per cent emission in any band.
% The millimetre continuum emission primarily originates from a nucleus and a southern star-forming region.  We detected H40$\alpha$ and H42$\alpha$ lines from the southern region but only detected H42$\alpha$ line in the nucleus, which can be explained by the lower signal to noise in the H40$\alpha$ spectra and/or the potentially blended H42$\alpha$ line emission with the $c$-C$_3$H$_2$ line.

%The continuum emission corrected using total infrared emission yields low SFRs, probably because of the heavily obscured region, while the continuum emission corrected using mid-infrared emission yields SFRs comparable to the SFR from ALMA data, which indicates a relatively hot dust within NGC~1808 compared to typical star-forming galaxies.

\end{abstract}

\begin{keywords}
galaxies: individual: NGC 1808 -- galaxies: starburst -- galaxies: star formation -- radio continuum: galaxies -- radio lines: galaxies.
\end{keywords}

%%%%%%%%%%%%%%%%%%%%%%%%%%%%%%%%%%%%%%%%%%%%%%%%%%

%%%%%%%%%%%%%%%%% BODY OF PAPER %%%%%%%%%%%%%%%%%%

\section{Introduction}\label{sec:intro}

%% Star formation in galaxies is a subject at the root of astrophysics. 
Extragalactic star formation rates (SFRs) are usually calculated from ultraviolet continuum emission, optical and near-infrared recombination lines (e.g., $\ha$), mid- and far-infrared continuum emission, and radio continuum emission. Each of these SFR tracers has different advantages and disadvantages \citep{Kennicutt2012}. The main advantage of ultraviolet continuum emission and optical and near-infrared recombination lines is that they can directly trace young stellar populations and are therefore more directly related to SFR than most other tracers. However, ultraviolet continuum emission and optical recombination lines are heavily affected by dust obscuration. Although dust obscuration has less of an influence on near-infrared recombination line emission, it is still an issue in dusty starburst galaxies \citep[e.g.,][]{Marconi2000}. Infrared dust continuum emission and radio continuum emission are unaffected by dust obscuration in general, but the chief disadvantage of both is that they cannot directly trace the young stellar populations. The infrared continuum emission is a tracer of bolometric stellar luminosity, and hence using it to calibrate the SFR may result in an overestimate if numerous older evolved stars are present. Since the radio continuum emission is a combination of free-free continuum emission and synchrotron radiation, the two must be separated from each other to accurately measure SFRs. In addition, the most widely used calibration of the conversion from radio continuum emission to SFR is not based on first principles but on the empirical radio-IR correlation, which introduces extra systematic uncertainties. 
% Besides, since cosmic rays that produce synchrotron radiation travel across tremendous distances through the interstellar medium (ISM), radio emission appears diffused in comparison to star formation on scales of $\sim$100~pc \citep{}.

SFRs can be measured by free–free continuum and recombination line emission at millimetre wavelengths as well. On the one hand, unlike infrared or radio continuum SFR metrics, millimetre free–free continuum and millimetre recombination line emission is produced by photoionized gas, so it can directly trace young stellar populations.  On the other hand, unlike ultraviolet, optical, and near-infrared SFR metrics, the millimetre continuum and recombination line emission are generally unaffected by dust attenuation.  Additionally, unlike recombination line emission at centimetre or longer wavelengths, which is generally affected by a mix of masing effects and opacity issues in the photoionized gas, such effects are negligble at millimetre wavelengths \citep{Gordon1990}. 

Previous works have rarely employed millimetre recombination line emission to measure SFRs because it is difficult to detect such faint line emission. Before the Atacama Large Millimeter/submillimeter Array (ALMA) went into operation, the extragalactic millimetre recombination line emission had only been detected in M82 \citep{Seaquist1994,Seaquist1996}, NGC 253 \citep{Puxley1997} and Arp 220 \citep{Anantharamaiah2000}. However, ALMA has been capable of detecting millimetre continuum and recombination line emission in more nearby infrared-luminous galaxies \citep{Scoville2013}, and ALMA detections of recombination line emission have been published for NGC 253 \citep{Bendo2015,Meier2015}, NGC 4945 \citep{Bendo2016}, NGC 5253 \citep{Bendo2017}, NGC 3256 \citep{Michiyama2020} and the Circinus Galaxy \citep{Wang2023}.  Additionally, potential recombination line emission has been detected from from Arp 220 \citep{Scoville2015}, although it is also noteworthy that no recombination line emission was detected from NGC 1068 \citep{Izumi2016}.  
The analyses of \citet{Bendo2015,Bendo2016,Bendo2017} included the comparisons of SFRs from free–free continuum and millimetre recombination line emission with other star formation tracers, which can provide us with opportunities to examine the fundamental assumptions behind various star formation metrics in specific situations.  For instance, the inconsistencies between SFRs from millimetre recombination line emission and mid-infrared continuum emission in NGC~4945 illustrate that the heavy dust obscuration in this galaxy strongly affects star formation metrics based on mid-infrared data \citep{Bendo2016}.  Meanwhile, in the metal-poor dwarf galaxy NGC~5253, the individual infrared bands yielded different SFRs because the shape of the spectral energy distribution (SED) of the galaxy differs significantly from what is found in standard spiral galaxies \citep{Bendo2017}.  Given how useful these studies have been, any additional measurements of SFRs from millimetre recombination line emission in particular but also from millimetre free-free continuum emission could be very useful for evaluating the accuracy of other SFR tracers and providing insights into the conditions under which these tracers cease to work accurately. 

% In this paper, we report on the ALMA detections of H40$\alpha$ (99.02 GHz) and H42$\alpha$ (85.69 GHz) recombination lines emission from the centre of NGC 1808. 
In this paper, we report on the ALMA detections of 85.69 and 99.02 GHz continuum emission and H42$\alpha$ and H40$\alpha$ lines emission from the centre of NGC 1808.  This object is a nearby (9.5$\pm$2.0~Mpc; \citealt{Tully2016}) barred spiral galaxy \citep{deVaucouleurs1991} with a starburst nucleus.  The galaxy is known for its peculiar hot spots in its dusty central region \citep{Morgan1958,Sersic1965} where there are supernova remnants \citep[e.g.,][]{Saikia1990,Forbes1992,Collison1994} and young star clusters \citep[e.g.,][]{Krabbe1994,Kotilainen1996,Tacconi-Garman1996,Galliano2005,Galliano2008}.  It is unclear whether the galaxy contains an active galactic nucleus (AGN) at the centre, but even if an AGN is present, it may be a relatively weak energy source \citep[e.g.,][]{Phillips1993,Krabbe2001,Jimenez-Bail2005,Dopita2015,Busch2017}.  \citet{Salak2016,Salak2017,Salak2018,Salak2019} and \citet{Audibert2021} have studied this galaxy using ALMA data, but their analysis has primarily focused on the molecular gas.  
%[ZZZ George to rewrite the following sentence.]  
The compact circumnuclear starburst in NGC 1808 has one of the highest infrared and radio fluxes in the nearby universe, so is an ideal object for calibrating SFRs using millimetre continuum and recombination line emission and other star formation metrics.
%% Sec 4.6, Nuclear activity (Busch 2017).
%% the analysis of the HCN/HCO$+$ and HCN/CS ratios (Audibert 2021).

% Outline adding
The outline of this paper is as follows. We present the description of data in Section \ref{sec:data}.  The multi-wavebands images and the ALMA spectra for the centre of NGC~1808 are presented in Section \ref{sec:images_spectra}.  We present the analysis of SEDs in Section \ref{sec:SED} and the derivation of electron temperatures in Section \ref{sec:ET}.  Section \ref{sec:SFR} presents the measurements and comparisons of SFRs from ALMA and other data.  A brief summary of our main results is given in Section \ref{sec:conclusion}.

\section{Data}\label{sec:data}

This project is focused on the ALMA observations in project 2016.1.00562.S (PI: G. J. Bendo) of the H40$\alpha$ (99.02~GHz) and H42$\alpha$ (85.69~GHz) recombination lines emission and the underlying free-free emission from NGC 1808. However, our analysis depends on accurately characterizing the SEDs at these frequencies so that we can remove the contribution of synchrotron and dust emission, so we also use archival ALMA Band 3, 6, and 7 data as well as the National Radio Astronomy Observatory (NRAO) Very Large Array \citep[VLA;][]{Condon1998} L-, C-, X-, U-band data to create an SED spanning from 1 to 360~GHz. Additional ultraviolet, $\ha$, and infrared data from online archives are used to calculate alternate SFRs that can be compared to those from ALMA.

\subsection{ALMA Data} \label{subsec:ALMA}
%{ALMA Observations and Data Reduction}

%[Describe the strategy for selecting data for the analysis]
%[We have this Cycle 4 data]
%[We also need to measure SED data, so we grabbed stuff from the archive]

\subsubsection{Observations}

The ALMA observations for project 2016.1.00562.S were performed on 2017 Jan 05 using the main (12 m) array.  The observations consist of a single pointing aimed at the centre of NGC 1808. The spectral windows, including the H40$\alpha$ and H42$\alpha$ lines, were centred on observed 98.69 and 85.40 GHz, respectively.  Both contained 1920 channels with a width of 976.6~kHz (3.0~km~s$^{-1}$ for H40$\alpha$ and 3.4~km~s$^{-1}$ for H42$\alpha$) and included both XX and YY polarizations.

% [Other archival data observations for continuum]
For the continuum imaging at other frequencies, we used all other available ALMA Band 3, 6, and 7 observations from both the 12 m and 7 m Array, including  not only data from project 2016.1.00562.S but also data from 2012.1.01004.S, 2013.1.00911.S, and 2017.1.00984.S. General information for these ALMA observations is listed in Table \ref{tab:ALMA_obs}. We did not include any ALMA total power observations in our analysis because those data cannot be used for continuum imaging. 

%[Table for the information of ALMA observations]
\begin{table*}
\centering
\begin{minipage}{0.85\textwidth}%{219mm}
\caption{General information of ALMA observations$^a$.}
\label{tab:ALMA_obs}
\begin{tabular}{@{}cccccccccc@{}}
    \hline
    \hline
      Band &
      Project &
      Array &
      Scheduling &
      $5\%$-$80\%$ of  &
      Angular &
      Maximum &
      Observed &
      Number \\
       &
      code &
       &
      Block &
      baselines &
      resolution$^c$ &
      recoverable &
      frequency &
      of \\
       &
       &
       &
       &
      range$^b$ &
       &
      scale$^c$ &
      range &
      channels \\
       &
       &
       &
       &
      (m) &
      (arcsec) &
      (arcsec) &
      (GHz) &
       \\
    \hline
      3 &
      2016.1.00562.S &
      12 m &
      NGC\_1808\_a\_03\_TM1 &
      31-182 &
      2.1 &
      21.3 &
      84.42-86.29 &
      1920 \\
       &
       &
       &
       &
       &
       &
       &
      86.29-88.17 &
      1920 \\
       &
       &
       &
       &
       &
       &
       &
      96.34-98.22 &
      1920 \\
       &
       &
       &
       &
       &
       &
       &
      98.22-100.09 &
      1920 \\
       &
      2012.1.01004.S &
      7 m &
      NGC1808\_Band3\_ACA &
      9-33 &
      9.7 &
      64.1 &
      100.00-101.92 &
      124 \\
       &
       &
       &
       &
       &
       &
       &
      101.89-103.81 &
      124 \\
       &
       &
       &
       &
       &
       &
       &
      111.95-113.87 &
      124 \\
       &
       &
       &
       &
       &
       &
       &
      113.90-115.90 &
      4080 \\
       &
       &
      12 m &
      NGC1808\_Band3\_12m & %X50a
      35-195 &
      1.7 &
      16.4 &
      99.96-101.95  &
      128 \\
       &
       &
       &
       &
       &
       &
       &
      101.86-103.84 &
      128 \\
       &
       &
       &
       &
       &
       &
       &
      111.91-113.90 &
      128 \\
       &
       &
       &
       &
       &
       &
       &
      113.96-115.84 &
      3840 \\
       &
      2013.1.00911.S &
      12 m &
      NGC1808\_a\_03\_TC & % X2be
      25-159 & 
      2.3 &
      26.7 &
      85.58-87.46 &
      1920 \\
       &
       &
       &
       & % X2be
       &
       &
       &
      87.48-89.35 &
      1920 \\
       &
       &
       &
       & % X2be
       &
       &
       &
      97.58-97.81 &
      240 \\
       &
       &
       &
       & % X2be
       &
       &
       &
      99.42-101.41 &
      128 \\
       &
       &
       &
      NGC1808\_a\_03\_TE & % X2bc 
      102-580 & 
      0.66 &
      6.4 &
      85.58-87.4 &
      1920 \\
       &
       &
       &
       & % X2bc 
       & 
       &
       &
      87.48-89.35 &
      1920 \\
       &
       &
       &
       & % X2bc 
       &
       &
       &
      97.58-97.81 &
      240 \\
       &
       &
       &
       & % X2bc 
       &
       &
       &
      99.42-101.41 &
      128 \\
    \hline
      6 &
      2017.1.00984.S &
      7 m &
      NGC1808\_a\_06\_7M & %Xa54
      9-33 &
      4.6 &
      28.9 &
      227.05-229.03 &
      128 \\
       &
       &
       &
       & %Xa54
       &
       &
       &
      229.15-231.14 &
      512 \\
       &
       &
       &
       & %Xa54
       &
       &
       &
      242.77-244.77 &
      512 \\
       &
       &
       &
       & %Xa54
       &
       &
       &
      244.55-246.53 &
      128 \\
       &
       &
       &
      NGC1808\_b\_06\_7M & %Xa5e
      9-31 &
      5.1 &
      30.2 &
      217.46-219.46 &
      512 \\
       &
       &
       &
       & %Xa5e
       &
       &
       &
      219.01-221.01 &
      512 \\
       &
       &
       &
       & %Xa5e
       &
       &
       &
      231.03-233.02 &
      512 \\
       &
       &
       &
       & %Xa5e
       &
       &
       &
      233.22-235.21 &
      128 \\
       &
       &
      12 m &
      NGC1808\_a\_06\_TM1 & %Xa52
      27-186 &
      4.8 &
      9.5 &
      227.05-229.03 &
      128 \\
       &
       &
       &
       & %Xa52
       &
       &
       &
      229.21-231.08 &
      480 \\
       &
       &
       &
       & %Xa52
       &
       &
       &
      242.83-244.70 &
      480 \\
       &
       &
       &
       & %Xa52
       &
       &
       &
      244.55-246.53 &
      128 \\
       &
       &
       &
      NGC1808\_b\_06\_TM1 & %Xa5c
      28-191 &
      5.1 &
      9.6 &
      217.52-219.40 &
      480 \\
       &
       &
       &
       & %Xa5c
       &
       &
       &
      219.07-220.94 &
      480 \\
       &
       &
       &
       & %Xa5c
       &
       &
       &
      231.09-232.96 &
      480 \\
       &
       &
       &
       & %Xa5c
       &
       &
       &
      233.22-235.21 &
      128 \\
\hline
      7 &
      2013.1.00911.S  &
      7 m &
      NGC1808\_a\_07\_7M & %X26cf, X758
      7-29 &
      3.5 &
      24.8 &
      341.93-343.85 &
      124 \\
       &
       &
       &
       & %X26cf, X758
       &
       &
       &
      343.70-345.69 &
      124 \\
       &    
       &
       &
       & %X26cf, X758
       &
       &
       &
      354.59-356.58 &
      510 \\
       &
       &
       &
       & %X26cf, X758
       &
       &
       &
      355.88-357.80 &
      510 \\
       &
       &
      12 m &
      NGC1808\_a\_07\_TE & % raw_Xd7e
      30-169 &
      0.60 &
      5.7 &
      341.90-343.89 &
      128 \\
       &
       &
       &
       & % raw_Xd7e
       &
       &
       &
      343.76-345.63 &
      480 \\
       &
       &
       &
       & % raw_Xd7e
       &
       &
       &
      354.65-356.52 &
      480 \\
       &
       &
       &
       & % raw_Xd7e
       &
       &
       &
      355.85-357.83 &
      128 \\
\hline
\end{tabular}
$^a$The information is retrieved from ALMA Science Archive (\url{http://almascience.org/asax/}). \\
$^b$These values are $5\%$ and $80\%$ of all projected baselines calculated from the $uv$ distributions.\\
$^c$These values are estimated based on the used baselines.
\end{minipage}
\end{table*}

\subsubsection{Data calibration and imaging}

The ALMA visibility data were reprocessed using the {\sc common astronomy software applications} ({\sc casa}; \citealt{CASA2022}).  The general recommendation for pipeline-calibrated data is to use the same version of {\sc casa} that was originally used to calibrate the data.  Therefore, for data from project 2016.1.00562.S, we used  {\sc casa} version 4.7.0, and for data from project 2017.1.00984.S, we used {\sc casa} version 5.1.1.
% Information about the specific {\sc CASA} version used to calibrate each dataset is listed in Table \ref{Table_ALMA_obs}.

The other archival data from projects 2012.1.01004.S and 2013.1.00911.S were not originally pipeline calibrated, so we manually calibrated these data using {\sc casa} version 6.2.1.  We first applied phase corrections based on water vapour radiometer measurements, amplitude corrections based on system temperatures, and antenna position corrections to the data.  Next, we visually inspected the data and flagged any noisy or anomalous visibility data.  After this, we derived and applied corrections to the bandpass, phase, and amplitude. According to the ALMA Proposers' Guide \citep{Privon2022}\footnote{\url{https://almascience.eso.org/documents-and-tools/cycle9/alma-proposers-guide}}, the typical flux calibration uncertainties are 5 per cent for Band 3 and 10 per cent for Band 6 and 7 data.

\begin{table}
\centering
\begin{minipage}{0.425\textwidth}%{219mm}
\caption{Information of ALMA continuum imaging.}
\label{tab:ALMA_continuum}
%\begin{threeparttable}
% spectral frequency, $uv$ ranges and the restoring beam sizes
    \begin{tabular}{@{}ccccc@{}}
    \hline
    \hline
      Band &
      Project &
      Observed &
      $uv$ range &
      Beam \\
       &
      code &
      frequency &
      for imaging &
      FWHM$^\star$\\
       &
       &
      (GHz) &
      (m) &
      ($\rm {arcsec^2}$)\\
    \hline
      3 &
      2016.1.00562.S &
      86.29 &
      34-442 & 
      3.4$\times$2.0 \\
       &
       &
      98.22 &
      30-230 &
      3.2$\times$2.2 \\
       &
      2012.1.01004.S  &
      101.90 &
      29-194 & 
      3.0$\times$2.3 \\
       &
       &
      113.86 &
      26-168 & 
      3.0$\times$2.4 \\
      %\cline{2-6}
      &
      2013.1.00911.S &
      87.43 &
      34-268 & 
      3.1$\times$2.3 \\
       &
       &
      100.46 &
      29-213 & 
      3.0$\times$2.3 \\
    \hline
      6 &
      2017.1.00984.S &
      219.23 & 
      13-84 & 
      2.8$\times$2.5 \\
       &
       &
      229.09 &
      13-79 &
      2.8$\times$2.5 \\
       &
       &
      233.12 & 
      13-79 & 
      2.8$\times$2.5 \\
       &
       &
      244.66 & 
      12-74 & 
      2.8$\times$2.5 \\
    \hline
      7 &
      2013.1.00911.S &
      343.80 & 
      9-50 & 
      2.7$\times$2.6 \\
       &
       &
      356.23 &
      8-48 &
      2.7$\times$2.6 \\
    \hline
    \end{tabular}
        $\star$ All final beam FWHM are 2.7 arcsec ($\sim$120 pc for a distance of 9.5 Mpc), calculated by $\sqrt{a\times b}$, where $a$ and $b$ are the length of the major and minor axis, respectively.
%\end{threeparttable}
% \begin{flushleft}$^\star$\end{flushleft}
\end{minipage}
\end{table}

These processed visibility data were converted into recombination lines and continuum image cubes using {\sc tclean} in {\sc casa} version 6.2.1.  To begin with, we created image cubes without any continuum subtraction to identify the line-free regions in the data.  Next, two continuum images were created in the spectral windows containing the H40$\alpha$ (99.02~GHz) and H42$\alpha$ (85.69~GHz) lines; these images will be used for comparison to the line emission.  Additionally, continuum images for the SED analysis were created based on every two adjacent spectral windows with comparable frequencies of the visibility data.  When creating these images, the $uv$ coverages were adjusted so that the images have the same full width at half-maximum (FWHM) of the restoring beam (2.7 arcsec) of the 85.69~GHz continuum data and the same maximum recoverable scale (12.7 arcsec) of the 357~GHz data.  The details of the continuum imaging are listed in Table \ref{tab:ALMA_continuum}.

After this, the H40$\alpha$ and H42$\alpha$ image cubes were constructed from the continuum-subtracted visibility data.  To ensure a signal-to-noise ratio greater than 5 for both lines, the image cube channels are set to be 15 times the width of the visibility channels, which results in an image cube channel width of 14.65~MHz, equivalent to a velocity width of 44~km~s$^{-1}$ for H40$\alpha$ and 51~km~s$^{-1}$ for H42$\alpha$.  % The FWHM of the reconstructed beams are 3.15 $\times$ 1.88 and 3.67 $\times$ 2.18 $\rm {arcsec^2}$ for the H40$\alpha$ and H42$\alpha$ images.  

When these continuum images and recombination lines image cubes were created, we selected the H{\"o}gbom cleaning algorithm because it is effective for imaging data with relatively low signal-to-noise ratios.  For the continuum images, we used natural weighting to recover extended emission and to maximize the signal-to-noise ratio.  For the image cubes covering the recombination lines, we set the {\sc weighting} parameter to {\sc briggsbwtaper}, which includes frequency-related adjustments across the bandwidth of the cube, and we set the {\sc robust} parameter to 2 to make the weighting similar to natural weighting.  Primary beam corrections were applied to all images and image cubes.  The pixel scales are set as 0.4 arcsec to subsample the beam effectively and the image dimensions are set as 320$\times$320 spaxels (128$\times$128 $\rm {arcsec^2}$, corresponding to $\sim$6~kpc at a distance of 9.5 Mpc) to cover the primary beam.

\subsection{VLA Data}\label{subsec:VLA}  %{Radio}

\begin{table}
\centering
\begin{minipage}{0.49\textwidth}
\caption{Information of VLA continuum imaging.}\label{tab:VLA_continuum}
    \begin{tabular}{@{}cccccc@{}}
    \hline
    \hline
      Band &
      Project &
      Observing &
      Observed &
      $uv$ range &
      Beam \\
       &
      ID &
      dates &
      frequency &
      for imaging &
      FWHM$^\star$ \\
       &
       &
       &
      (GHz) &
      (m) &
      ($\rm {arcsec^2}$)\\
    \hline
      L &
      AV116 &
      1985 Feb 21 &
      1.51 &
      1938-25000 & 
      5.0$\times$1.4\\
      \multirow{2}{*}{C} &
      AU25 &
      1986 Jun 21 &
      4.89 &
      \multirow{2}{*}{598-3980} & 
      \multirow{2}{*}{3.0$\times$2.3} \\
       &
      AB1020 &
      2002 Jan 28 &
      4.89 &
       & 
       \\
      \multirow{2}{*}{X} &
      AW278 &
      1991 Sep 01 &
      8.46 &
      \multirow{2}{*}{345-2085} & 
      \multirow{2}{*}{2.8$\times$2.5} \\
       & 
      AB1020 &
      2002 Jan 28 &
      8.49 &
       & 
       \\
      % U & 
      % AU25 &
      % 1986 Jun 21 &
      % 14.9 &
      % 210-1215 & 
      % 3.5$\times$2.0 \\
    \hline
    \end{tabular}
    $\star$ All final beam FWHM are 2.7 arcsec, same as Table \ref{tab:ALMA_continuum}
\end{minipage}
\end{table}

The radio visibility data used in this work, which span from 1.5 to 8.5~GHz, were downloaded from the NRAO VLA Archive Survey \citep[NVAS\footnote{\url{http://www.vla.nrao.edu/astro/nvas/}};][]{Crossley2007,Crossley2008}, which processed the VLA raw visibility data from 1976 to 2006 with {\sc Astronomical Image Processing System} VLA pipeline to make the archive data easily accessible to the astronomical community.  For our SED analysis, we only used the calibrated visibility data with $uv$ ranges that allowed us to create images with beam FHWMs and maximum recoverable scales that are comparable to our ALMA continuum images.  It is worth noting that NVAS also provides visibility data at 14.9 GHz, but its $uv$ coverage cannot be adjusted to have the same maximum recoverable scale as that of ALMA images, so we did not include these data. 

The data from the NVAS were not in a format that could be read by {\sc tclean}, so we used {\sc clean} in {\sc casa} version 5.1.1 to image these data.  We again used the H{\"o}gbom algorithm and natural weighting.  The final VLA images have pixel sizes of 0.4 arcsec and dimensions of 128$\times$128 arcsec$^2$ to match the ALMA images.  The VLA 1.51 GHz image has an astrometry coordinate issue that we corrected using the world coordinate systems from the ALMA and other VLA images.  Based on the information from the VLA Observational Status Summary 2023A\footnote{\url{https://science.nrao.edu/facilities/vla/docs/manuals/oss/performance/fdscale}} the flux scaling uncertainties are 5 per cent for the L-band data and 15 per cent for the C-, X-, and U-band data.  General information for the VLA continuum images is shown in Table \ref{tab:VLA_continuum}.

\subsection{Infrared Data}

\begin{table*}
\centering
\begin{minipage}{0.84\textwidth} %{176mm}
\caption{Technical information for infrared images.}
\label{tab:IR}
\begin{tabular}{@{}cccccccl@{}}
\hline
\hline
Telescope &
  Instrument &
  Wavelength &
  Frequency &
  Beam &
  Flux &
  Pixel &
  References for\\
   &
   &
   &
   &
  FWHM &
  calibration &
  scale &
  technical information\\
&
  &
  ($\mu$m) &
  (GHz) &
  (arcsec) &
  uncertainty &
  (arcsec) &
  \\
\hline
WISE &
  &
  22 &
  13636 &
  17$^a$ &
  2\% &
  1.375 &
  \citet{Cutri2013}$^b$\\
{\it Herschel} &
  PACS &
  70 &
  4286 &
  5.6 &
  6\% &
  1.6 &
  \citet{Exter2019}$^c$ \\
{\it Herschel} &
  PACS &
  100 &
  3000 &
  6.8 &
  6\% &
  1.6 &
  \citet{Exter2019}$^c$ \\
{\it Herschel} &
  PACS &
  160 &
  1875 &
  10.7 &
  8\% &
  3.2 &
  \citet{Exter2019}$^c$ \\
{\it Herschel} &
  SPIRE &
  250 &
  1200 &
  18.4 &
  5.5\% &
  6 &
  \citet{Bendo2013},\citet{Valtchanov2018}$^d$ \\
\hline
\end{tabular}
$^a$ While the original beam FWHM is 12~arcsec, the beam FWHM in the final atlas tiles is 17~arcsec.\\
$^b$ This reference is available from \url{http://wise2.ipac.caltech.edu/docs/release/allwise/expsup/index.html} .\\
%%$^a$ The observed beam FWHM is 12~arcsec, although the FWHM in the final atlas tiles is 17~arcsec.\\
%$^c$ This value includes absolute uncertainty, 5\%, and relative uncertainty 1\%.\\
%$^d$ This value includes absolute uncertainty, 5\%, and relative uncertainty 3\%.\\
%$^e$ This value includes absolute uncertainty, 4\%, and relative uncertainty 1.5\% \citep{Bendo2013}.\\
$^c$ This reference is available from %
\url{https://www.cosmos.esa.int/web/herschel/legacy-documentation-pacs} .\\
% \url{https://www.cosmos.esa.int/documents/12133/996891/PACS+Explanatory+Supplement} . \\
% \url{http://herschel.esac.esa.int/Docs/PACS/pdf/pacs\_om.pdf} .\\\url{http://herschel.esac.esa.int/Docs/PACS/pdf/pacs\_om.pdf} .\\
% $^d$ These SPIRE beam FWHM apply to the timeline data, which was used for photometry measurements.  The beam appears broader in the map data.\\
$^d$ This reference is available from 
\url{https://www.cosmos.esa.int/web/herschel/legacy-documentation-spire} .\\
% \url{https://www.cosmos.esa.int/documents/12133/1035800/The+Herschel+Explanatory+Supplement\%2C\%20Volume+IV+-+THE+SPECTRAL+AND+PHOTOMETRIC+IMAGING+RECEIVER+\%28SPIRE\%29} .
% \url{http://herschel.esac.esa.int/Docs/SPIRE/spire\_handbook.pdf} .\\\url{http://herschel.esac.esa.int/Docs/SPIRE/spire\_handbook.pdf} .\\
\end{minipage}
\end{table*}

The infrared-based SFRs used in this work were calculated using 22~$\mu$m data from the Wide-field Infrared Survey Explorer \citep[WISE; ][]{Wright2010} and 70-250~$\mu$m archival data from the {\it Herschel} Space Observatory \citep{Pilbratt2010}.  The technical details for the infrared data used in this work are listed in Table \ref{tab:IR}.  

The {\it WISE} 22 $\mu$m image used in our analysis is the atlas tile 0776m379\_ac51 produced by the WISE Science Data System version 6.0.  This image is a 1$^\circ$.56 square and includes the entire galaxy.  While {\it Spitzer} Space Telescope \citep{Werner2004} 24~$\mu$m data are also available for this galaxy, those data are saturated, which is why we used the {\it WISE} 22~$\mu$m data.  

% \footnote{https://wise2.ipac.caltech.edu/docs/release/allsky/expsup/sec4\_4h.html}
% https://wise2.ipac.caltech.edu/docs/release/allwise/expsup/sec4_3a.html

The 70, 100, and 160~$\mu$m images were acquired with the Photoconductor Array Camera and Spectrometer \citep[PACS; ][]{Poglitsch2010} on {\it Herschel} in four observations (Observation IDs 1342204260, 1342204261, 1342204262 and 1342204263).  The former two observations produced 70 and 160~$\mu$m images while the latter two observations produced 100 and 160~$\mu$m images. Each pair of observations is a set of scan maps made using a scan speed of 20 arcsec s$^{-1}$ in orthogonal directions to cover 12$\times$12~arcmin$^2$ regions.  The data were processed using the standard SPIRE pipeline in HIPE version v14.2.0.  

%\subsubsection{Herschel 250 $\mu$m data}

The 250~$\mu$m image was acquired with the Spectral and Photometric Imaging REceiver \citep[SPIRE; ][]{Griffin2010} instrument on {\it Herschel} in observation 1342203633.  The observation is a large scan map of a 30$\times$15~arcmin$^2$ region, which was scanned in two orthogonal directions at a speed of 30 arcsec s$^{-1}$.  The data were processed using the standard SPIRE pipeline in HIPE version v14.1.0 and calibration tree spire\_cal\_14\_3.

\subsection{H$\alpha$ Data} \label{subsec:ha}
% http://archive.eso.org/cms/eso-data-access-policy.html

The $\ha$ data was acquired with the Multi Unit Spectroscopic Explorer \citep[MUSE;][]{Bacon2010} on the Very Large Telescope.  The observations of NGC 1808 were taken in the Wide Field Mode in four observations (Program ID 0102.B-0617(A); \citealt{Perna2021}).  We used the data from the ESO Phase 3 Data Release\footnote{Based on data obtained from the ESO Science Archive Facility with DOI(s): \url{https://doi.eso.org/10.18727/archive/42} .} \citep[labelled ``MUSE-DEEP'']{Hanuschik2019} which are reduced with MUSE pipeline {\sc muse-2.6.2} \citep{Weilbacher2020} by removing instrument signature and combining all products from that step into the data cube available from the archive.  The image covers a field of view of 1$\times$1 arcmin$^2$ and has a sampling of 0.2$\times$0.2 arcsec$^2$ and a FWHM of 0.89 arcsec.  The spectra have a resolution of $\sim$2.6 \AA{} at 6500 \AA{} and a sampling of 1.25 \AA{} across the wavelength range from 4750 to 9350 \AA{}.  % Absolute fluxes of emission lines: Although Castro et al. (2018) find a few large deviations in the measured fluxes compared to the literature of up to 28%, more typical values range from 2 – 7%. They convincingly explain that most of the differences come from errors in the reported positions of the slits in the comparison work. 

Before we created the integrated $\ha$ image of the data, we need to correct missing data in the image cubes from 2 pixels near the galactic centre and 13 pixels offset by $\sim$3 arcsec that are saturated at $\sim 6563$\AA{}.  Since the $\ha/\hb$ ratios should not vary significantly in such small spatially adjacent pixels (9~pc), we assumed these saturated spectra pixels have the same ratios as the average $\ha/\hb$ ratios from spatially adjacent pixels.  We therefore used $\hb$ flux densities measured within the $\ha$ saturated spectra pixels multiplied by average $\ha/\hb$ ratios of spatially adjacent pixels to fill in the missing $\ha$ flux densities.  Our results based on $\ha$ data remain unchanged even if we directly use the maximum measured $\ha$ value to fill in the missing data because the number of $\ha$ saturated pixels is only a small fraction (15/7794) of the central 22 arcsec diameter region. % used in the following analysis.  

The final image of the integrated $\ha$ flux was created by first subtracting the continuum emission and then integrating the image cube data over the observed wavelength range at $\sim 6563$\AA.  To do the continuum subtraction, we first created a spectrum integrated over the central 22 arcsec (1~kpc) diameter region.  Next, we used the Penalized PiXel-Fitting (pPXF; \citealt{Cappellari2017}) package to perform a joint fitting with the Gaussian emission lines templates and stellar population models based on the Medium resolution Isaac Newton Library of Empirical Spectra \citep{Vazdekis2010} to the spectrum after it was corrected for Galactic extinction given by \citep{Schlegel1998} and shifted it to the rest frame.  Following this, the continuum emission for each image cube was determined by assuming the same continuum shape as the average continuum emission within the central 22 arcsec (1~kpc) diameter region.  After we subtracted the continuum emission for each image cube, we created the final image of the integrated $\ha$ flux by integrating the image cube data over the observed wavelength range from 6556 to 6569 \AA{},  which covers the $\ha$ line over a velocity width of $600 \kms$ while excluding the adjacent [N{\sc II}] lines.

% The $\ha$ image of this galaxy can also be acquired with the Survey for Ionization in Neutral Gas Galaxies \citep[SINGG;][]{Meurer2006} as well. The photometric calibration of the image yields an flux accuracy of $\approx 4\%$. We adopt their measurements of the absorption of $\ha$ by Galactic foreground dust, 0.8 mag, the $\nii$ contamination, $13.79\%$, and the uniform correction factor for the effect of underlying stellar absorption used in SINGG sample, $4\%$. % Before analysis, the $\ha$ flux are supposed to be corrected for the effects of Galactic foreground, $\nii$ contamination, and underlying $\ha$ absorption. 

\subsection{Ultraviolet Data} \label{subsec:UV}

% ; $\sim$1528$\AA$
% ; $\sim$2310$\AA$
The far-ultraviolet (FUV) and near-ultraviolet (NUV) images were acquired with the Galaxy Evolution Explorer ({\it GALEX}; \citealt{Martin2005}) in Observation ID 2504214835575652352 as part of the GALEX ultraviolet atlas of nearby galaxies \citep{GildePaz2007}.  The images cover a circular field of view of 1.2 degrees and have pixel scales of 1.5 arcsec.  The FWHM of the FUV and NUV images are 4.2 and 5.3 arcsec, respectively.  The photometric calibration yields flux accuracies of 5 per cent and 3 per cent for the FUV and NUV images \citep{Morrissey2007}.  Using the extinction coefficients for the GALEX bands given by \citet{Yuan2013} and the Galactic extinction given by \citet{Schlafly2011}, we derived FUV and NUV foreground extinctions of 0.153 and 0.227 mag.
% \citet{Marino2010} measure 4.88E-03, 9.64E-03 in 6.5$\times$3.9~arcmin regions
% ebv 0.027, egr 0.031

\section{Images and Spectra} \label{sec:images_spectra}

\subsection{Multi-waveband images} \label{subsec:images}

\begin{figure*}
  \centering
  \includegraphics[width=0.75\textwidth]{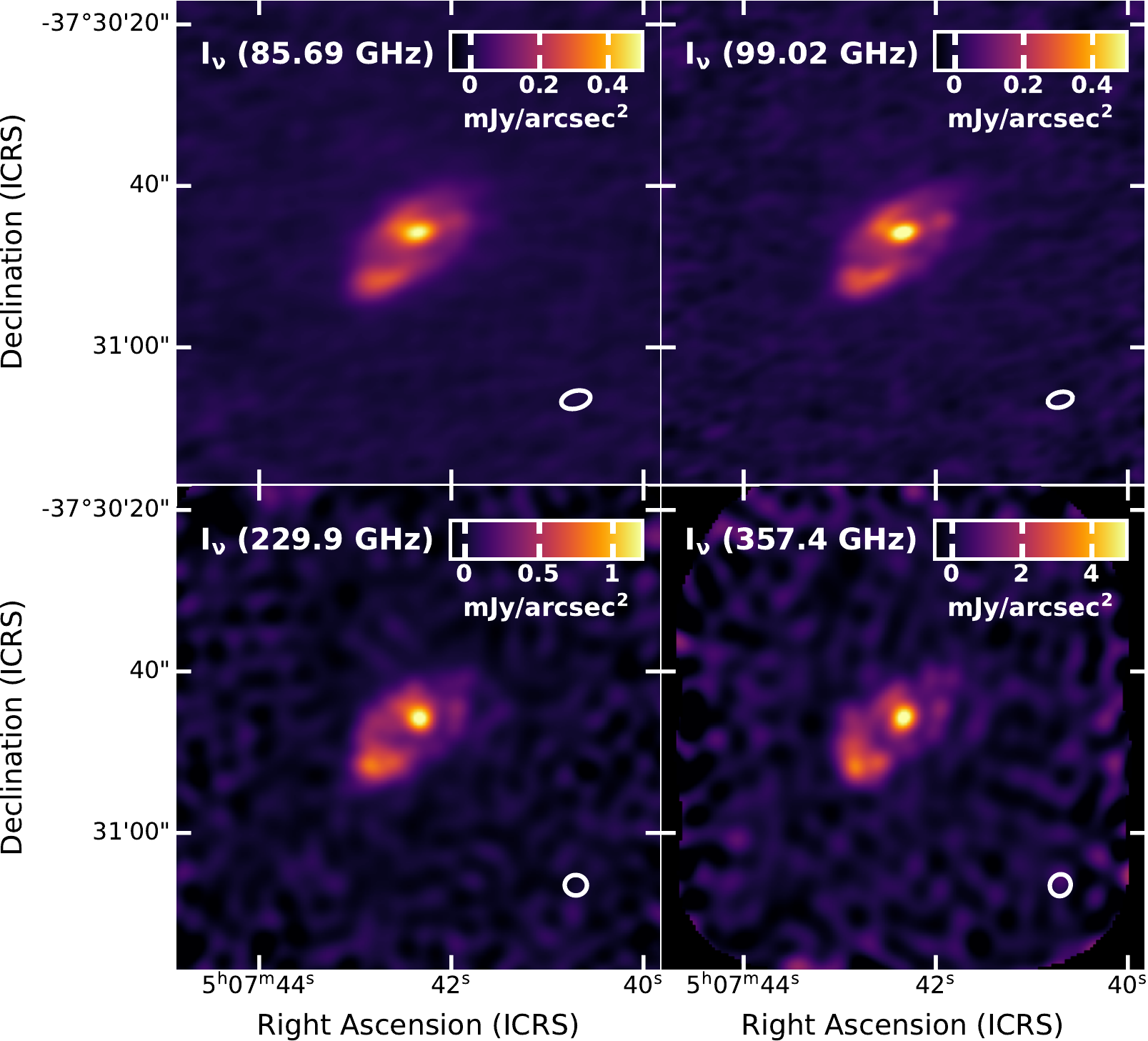} 
  \caption{Images of the ALMA 85.69~GHz, 99.02~GHz, 229.9~GHz and 357.4~GHz continuum in the central 60 arcsec of NGC 1808.  In each panel, the restoring beam is illustrated as the white open ellipse at the lower-right side.}
  % The green open star in the first panel marks the galactic position from NED.
  \label{fig:Image_ALMA}
\end{figure*}

\begin{figure*}
  \centering
  \includegraphics[width=0.95\textwidth]{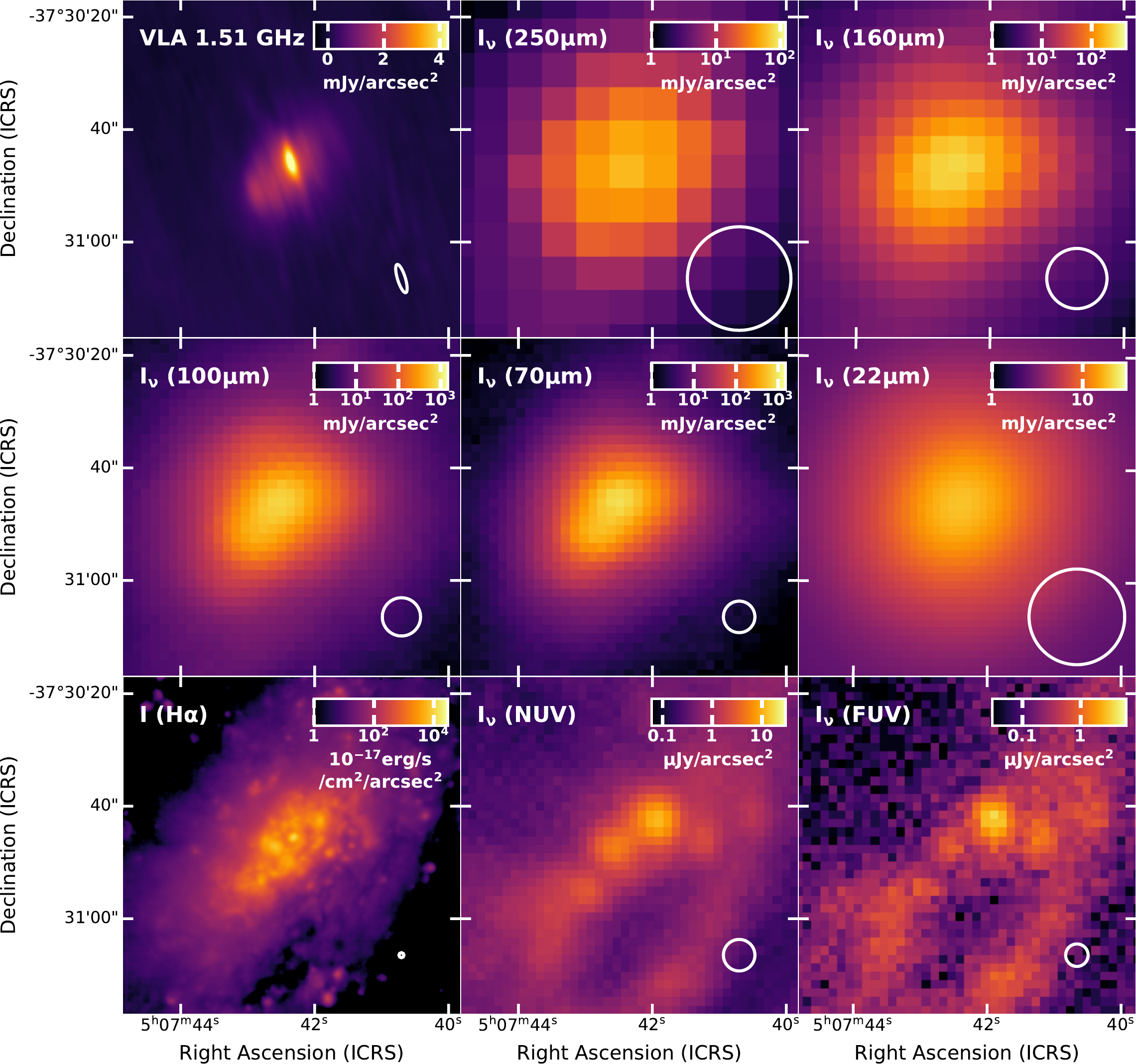} 
  \caption{Multiwaveband images of the VLA 1.51 GHz continuum, {\it Herschel} 250 $\mu$m, 160 $\mu$m, 100 $\mu$m and 70 $\mu$m continuum, {\it WISE} 22 $\mu$m continuum, MUSE $\ha$ line, and GALEX NUV and FUV continuum emission in the central 60 arcsec of NGC 1808.}
  % The green open star in the first panel marks the galactic position from NED.
  \label{fig:Image_other}
\end{figure*}
% \footnotemark
% \footnotetext{https://ned.ipac.caltech.edu/}

% The images of multi-waveband observations in the central 60~arcsec of NGC~1808 are displayed in the Figure~\ref{fig:Image}.  It is clear that the morphologies differ among these bands.  

% Both 99.02~GHz and 85.69~GHz continuum emission are mainly produced by photoionized gas (see more in Section \ref{sec:SED}) and unaffected by dust obscuration.  These images show that the star formation is mainly concentrated in the nucleus, but several other star-forming regions (e.g., the southeastern one) produce continuum emission detectable by ALMA.  

The images of ALMA observations in the central 60~arcsec of NGC~1808 are displayed in the Figure~\ref{fig:Image_ALMA}.  The 85.69~GHz and 99.02~GHz continuum emission are mainly produced by photoionized gas (see more in Section \ref{sec:SED}) and unaffected by dust obscuration.  These two images show that the star formation is mainly concentrated in the nucleus, but several other star-forming regions (e.g., the southeastern one) also produce continuum emission detectable by ALMA.  The 229.9~GHz and 357.4~GHz continuum emission are dominated by thermal dust emission (Section \ref{sec:SED}), but they also trace structures that are similar to what is seen in the 85.69~GHz and 99.02~GHz images at comparable spatial resolution.  This is probably because the dust within the central NGC~1808 absorbed the energy from the star-forming regions that are traced by the ALMA Band 3 continuum emission.

The images of other star formation tracers within the centre of NGC~1808 are displayed in the Figure~\ref{fig:Image_other}.  The morphologies differ among these bands.  

The nuclear and extended emission is also visible in the 1.51 GHz continuum images.  Although the orientation of the beam in the 1.51 GHz image differs from the beam of the ALMA images, the structures in the ALMA and VLA data are largely identical.

The beam sizes of the infrared images shown in Figure~\ref{fig:Image_other} are much larger than those of the other images used in this analysis.  The beams are so large in the 22 and 250$\mu$m images that the central structure is unresolved.  In the 70, 100, and 160~$\mu$m data, however, extended structures can be seen that are similar to what is seen in the ALMA data, although these structures are more poorly resolved in these infrared data.

% The images of infrared continuum emission are shown in the second row and the left and middle panels of the third row of Figure \ref{fig:Image}.  Compared with the radio and millimetre continuum emission, the infrared continuum emission looks more diffuse and continuous.  Furthermore, the coverage of morphology traced by the far-infrared continuum emission is more extended with wavelength increasing from 70 $\mu$m to 250 $\mu$m.  Rather than the difference in the spatial resolution, this variation of the far-infrared continuum emission with wavelength may arise from the difference in the distribution of warm dust traced by emission at 70-160 $\mu$m and cold dust traced by emission at wavelength greater than 160 $\mu$m \citep[e.g.,][]{Boquien2011}.  In addition, aside from the effect of the large beam FWHM, the circular morphology shown in the 22 $\mu$m image may also reflect the diffuse distribution of thermal continuum emission from small dust grains heated by intense radiation fields.

The last three panels of Figure \ref{fig:Image_other} show images of the MUSE $\ha$ line emission and the GALEX ultraviolet continuum emission, which can directly trace the young stellar populations, similar to free-free emission.  However, their morphologies appear different from those of the millimetre continuum emission because they are heavily affected by dust obscuration in the centre of the galaxy.  This is indicated by the $\ha/\hb$ flux ratio of $\sim$7 and FUV$-$NUV colour of $\sim$1.7 within the central 1 kpc region, both of which are significantly higher than what is found in nearby star-forming galaxies \citep{Kennicutt2009, Hao2011}.  Compared to the millimetre continuum emission, the $\ha$ line emission traces a more extended disc that includes the nucleus and multiple sources of comparable brightness, while the ultraviolet continuum emission traces a few unresolved bright sources and an arc structure.  The ultraviolet continuum emission from the galactic center is heavily obscured, but strong emission is seen from two neighbouring sources.  The brightest ultraviolet source, which is located $\approx 4$ arcsec (180 pc) northwest from where the millimetre continuum emission peaks, has a counterpart in $\ha$ line emission but not in millimetre continuum emission.  This location appears to be relatively unobscured, which would explain why the region produces strong ultraviolet emission, but at the same time, the SFR in the region may be relatively low, which would explain the relatively weak millimetre emission.  The millimetre-bright southern star-forming region corresponds to a relatively modest, clumpy $\ha$ source but not to any ultraviolet source.  

\subsection{ALMA spectra} \label{subsec:spectra}

\begin{figure}
  \centering
  \includegraphics[width=0.425\textwidth]{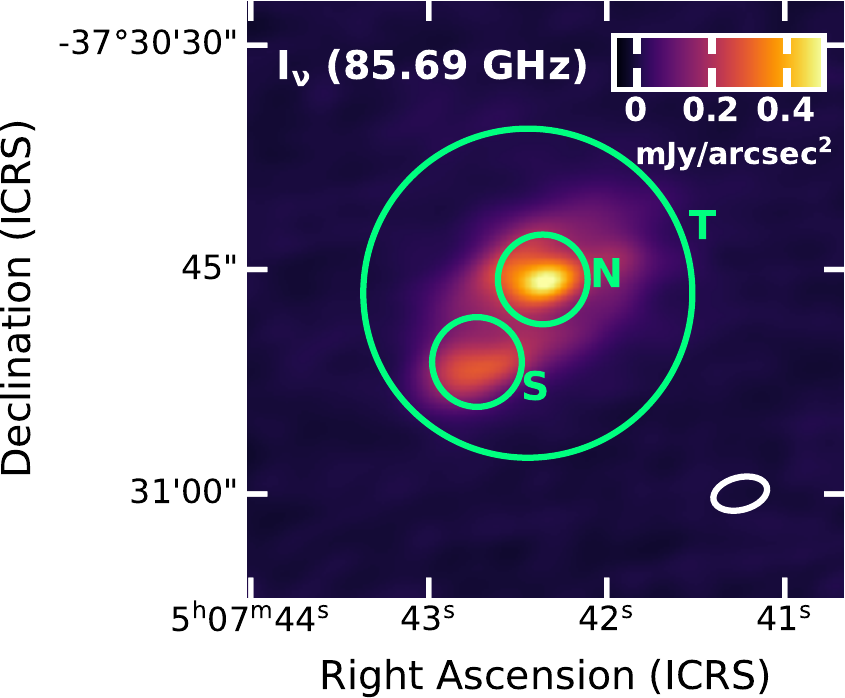} 
  \caption{The image of 85.69~GHz continuum emission in the central 40~arcsec of NGC 1808. The solid green circles identify the locations where we extracted spectra, including the total (T), nucleus (N) and south (S) apertures with diameters of 22~arcsec (1~kpc) , 6~arcsec (280~pc) and 6~arcsec (280~pc). The restoring beam is shown as the white open ellipse in the lower-right corner.}
  % In the middle and right panels, the radio velocities in the Kinematic Local Standard of Rest (LSRK) frame are texted at the upper-left side.}
  \label{fig:Image_Aper}
\end{figure}

\begin{figure*}
  \centering
  \includegraphics[height=0.835\textwidth]{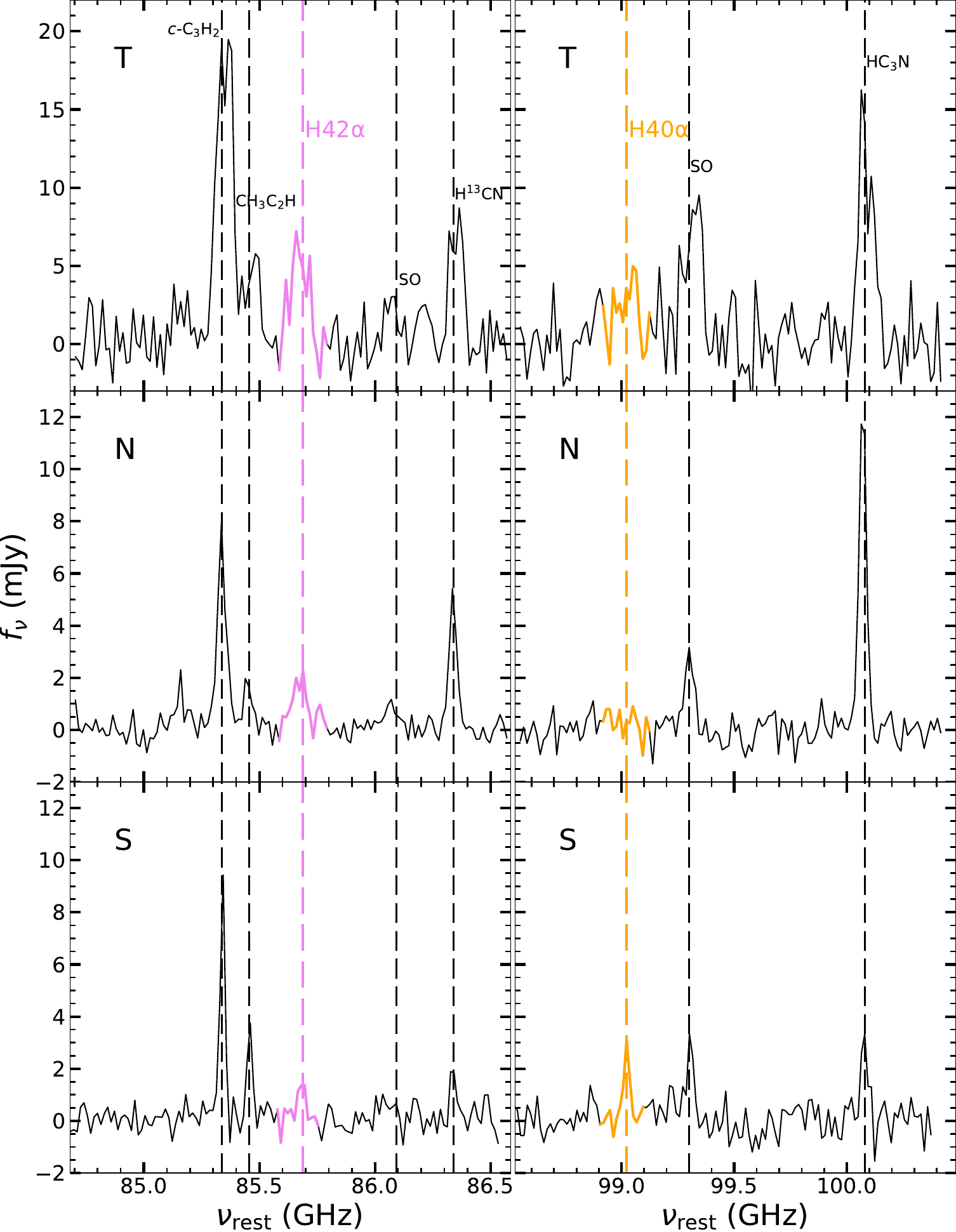} 
  \includegraphics[height=0.835\textwidth]{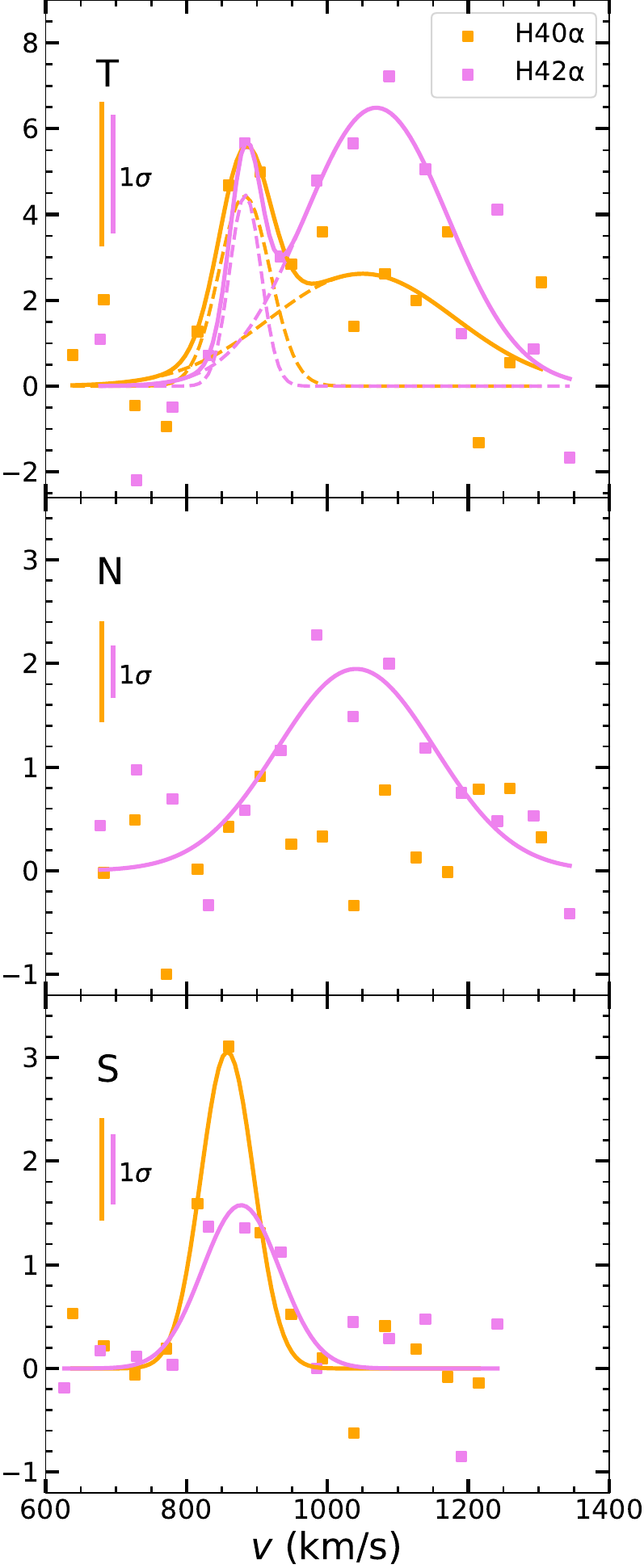}
  \caption{Spectra for the regions illustrated in Figure~\ref{fig:Image_Aper}. 
  The left and middle panels show the spectra measured within the entire spectral windows.  The frequencies have been separately shifted to the rest frame using the velocities of 990, 990 and 860 $\kms$ for the T, N and S regions (see text for the measurement).  The dashed lines mark the locations of the H42$\alpha$ (purple) and H40$\alpha$ (orange) line emission.  The panels on the right show the spectra in detail at the locations of the H40$\alpha$ and H42$\alpha$ lines.  The solid lines for the T region (top) show the best fit with two Gaussian functions, and the dotted lines show the individual Gaussian components in these fits.  The solid lines for the N region (middle) and S region (bottom) show the profiles fit with single Gaussian functions for the recombination lines.  The parameters for all of these functions are listed in Table \ref{tab:Gaussian_RL}.}
  \label{fig:Spec_RL}
\end{figure*}
% The frequencies have been separately shifted to the rest frame using the velocities of 970, 970 and 870 $\kms$ for the T, N and S regions, which are the average velocities of H40$\alpha$ and H42$\alpha$ lines measured from the T and S regions (Table \ref{tab:Gaussian_RL}). 

Figure \ref{fig:Image_Aper} shows the locations where we extracted spectra for analysis of the recombination lines.  Two measurement apertures with diameters of 6~arcsec (280~pc) were centred on the nucleus (N) and the bright continuum source to the south (S).  A third measurement aperture (T) with a diameter of 22~arcsec (1~kpc) was used to measure the total detectable emission from the central region.

Figure \ref{fig:Spec_RL} shows the spectra of H40$\alpha$ and H42$\alpha$ lines extracted from these three regions.  These spectral also reveal the presence of multiple other molecular lines, including $c$-C$_3$H$_2$, CH$_3$C$_2$H, SO and H$^{13}$CN in the lower frequency spectral window and SO and HC$_3$N in the higher frequency spectral window. %, but we are primarily focused on the hydrogen recombination line emission.
We used single Gaussian functions to fit the CH$_3$C$_2$H and H$^{13}$CN lines, which are the most significant line emission in their spectra windows, and obtained average relativistic velocities of $990\pm10$ $\kms$ and $860\pm10$ $\kms$ for the N and S regions, which we used to shift the frequencies to the rest frame.  We also adopted 990 $\kms$ as the rest frequency for the T region.
% The absence of any broad line emission indicates that the emission originates mainly from photoionized gas associated with the central starburst, rather than from a possible AGN.  

Significant H42$\alpha$ line emission is detected in both the N and S regions, and the emission can be fit by single Gaussian functions.  The mean relativistic velocity ($v_0$) measured in the S region is $\approx$160 $\kms$ lower that that of the N region.  This difference produces a double-peaked line profile in the T region, which we have fitted with two Gaussian functions. 

H40$\alpha$ line emission is detected in the S region, and this emission can be described by a single Gaussian function as well. However, no significant H40$\alpha$ line emission is detected in the N region; the signal is lower than the rms noise.  We also fit the profile of the H40$\alpha$ line in the T region using two Gaussian functions. The parameters for these Gaussian functions are listed in Table \ref{tab:Gaussian_RL}.

\begin{table}
\centering
\begin{minipage}{0.5\textwidth}
\caption{Parameters for Gaussian functions fit to the H42$\alpha$ and H40$\alpha$ lines.}\label{tab:Gaussian_RL}
    \begin{tabular}{@{}cccccc@{}}
    \hline
    \hline
       &
       &
      \multicolumn{2}{c}{H42$\alpha$} &
      \multicolumn{2}{c}{H40$\alpha$} 
      \\
      Region &
      Component &
      $v_0$ &
      FWHM &
      $v_0$ &
      FWHM %
       \\
       &
       &
      ($\kms$) &
      ($\kms$) &
      ($\kms$) &
      ($\kms$)\\
    \hline
      \multirow{2}{*}{Total} &
      lower $v_0$ &
      880 $\pm$ 40 & 
      50 $\pm$ 50 &
      880 $\pm$ 20 &
      90 $\pm$ 60 \\
       &
      higher $v_0$ &
      1070 $\pm$ 30 &
      240 $\pm$ 70 &
      1050 $\pm$ 90 &
      300 $\pm$ 190 \\
      Nucleus &
      single &
      1040 $\pm$ 20 &
      260 $\pm$ 50 &
       / &
       / \\
      South &
      single &
      880 $\pm$ 10 &
      130 $\pm$ 30 & 
      860 $\pm$ 10 &
      90 $\pm$ 10 \\
    \hline
    \end{tabular}
\end{minipage}
\end{table}

It is strange that the H40$\alpha$ line emission is detected in the S region but not the N region while the H42$\alpha$ line emission is detected in both.  Notably, in the N region, the noise in the spectrum of the H40$\alpha$ line is approximately twice as larger as the noise in the spectrum of the H42$\alpha$ line.  It may be possible the H40$\alpha$ line is not detected in this region simply because of the higher noise levels and the relatively low line flux.  However, other scenarios could explain why H42$\alpha$ emission but not H40$\alpha$ emission is detected.  We discuss these scenarios below.  

The first scenario is that the H40$\alpha$ line emission cannot be detected in the N region because of calibration errors or other related technical issues.  However, the H40$\alpha$ emission is still detected in the S region with an integrated signal-to-noise level that is comparable to the H42$\alpha$ emission, which would be difficult to achieve if the H40$\alpha$ line specifically was affected by calibration issues.  It therefore seems unlikely that any observing or data processing issues affected the H40$\alpha$ line from the nuclear (N) region specifically.  % If H40$\alpha$ and H42$\alpha$ line emission in the N region were similar to that in the S region, ALMA should have detected it.   

The second scenario is that the H42$\alpha$ emission from the nucleus is unusually enhanced.  The analysis from \citet{Gordon1990} indicates that the H40$\alpha$ and H42$\alpha$ emission should be spontaneous emission and that nonlinear effects should only gradually appear in higher order lines starting with H50$\alpha$.  With the exception of a claim by \citet{Baez2018} that the H26$\alpha$ emission exhibits masing from a subregion in NGC~253 but H30$\alpha$ does not (in a location on the edge of a much brighter source, which makes the analysis difficult), no one has reported observations of strongly enhanced emission in specific individual recombination lines in any extragalactic sources.  It would therefore seem unlikely that such an unusual jump in flux from H40$\alpha$ to H42$\alpha$ emission in the nucleus of NGC~1808 could be explained by line-specific nonlinear effects.
 
%The second possibility is the hydrogen recombination line maser emission in the H42$\alpha$ transition toward the N region.  The only extragalactic hydrogen recombination line maser emission was reported by \citet{Baez2018}, who identified the maser in the H26$\alpha$ transition next to the galactic nuclear of NGC~253.  However, the hydrogen maser is less likely to happen in the transitions where the principle quantum number is raised rather than lowered.  Another related conjecture is the overcooling proposed by \citet{Strelnitski1996}, which may occur for hydrogen recombination lines with $n \gtsim 20$ in $n_e \lesssim 10^5\rm{cm}^{-3}$.  Nevertheless, no strong absorption characteristic of the H40$\alpha$ line has been observed and no overcooling has been detected in other galaxies.  Therefore, even though we cannot completely rule out these possibilities based on the current data, it is worth noting that both scenarios hardly ever occur in extragalactic universe.  Future more sensitive millimetre observations could test these conjectures.   
% \citet{Baez2018} used the recombination line spectral index (RLSI) of $\gtsim 2$, which are determined from the ratio of intensities of H30$\alpha$ and H26$\alpha$ lines integrating over velocity.  The RLSI derived from H42$\alpha$ and H40$\alpha$ lines in the N region of NGC~1808 is of $\sim 4$, which is even higher than that found in NGC~253. 
% This scenario can explain the lack of H40$\alpha$ line emission in the galactic nucleus where H42$\alpha$ line emission is detected.  

\begin{figure}
  \centering
  \includegraphics[width=0.425\textwidth]{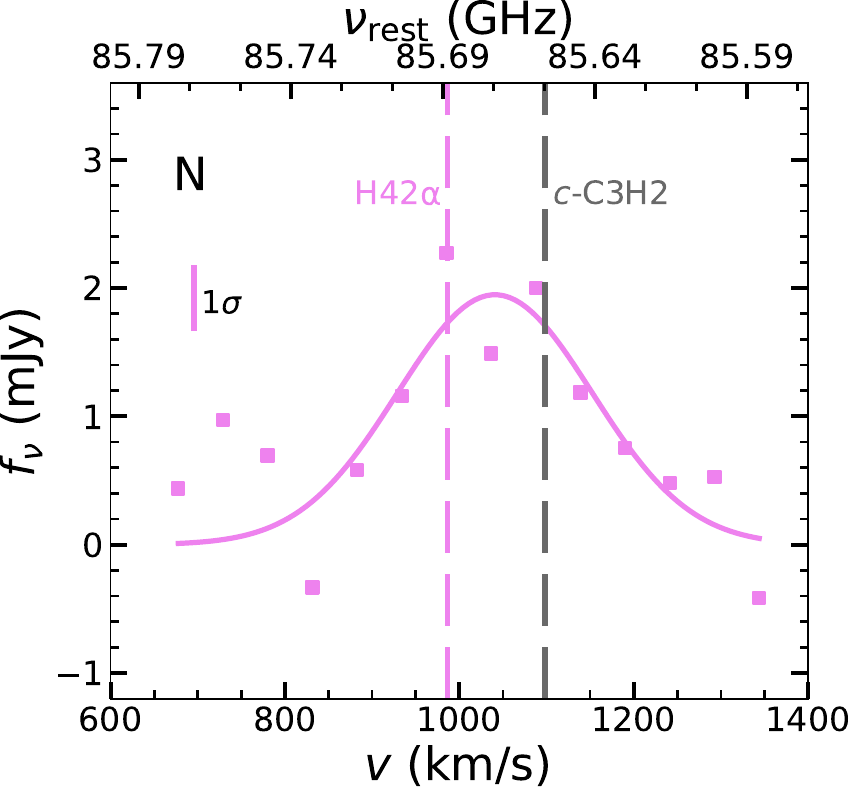} 
  \caption{Spectra of the H42$\alpha$ line for the N region illustrated in Figure \ref{fig:Image_Aper}.  The gray dashed line shows the location of the $c$-C$_3$H$_2$ ($4_{32}$$\rightarrow$$4_{23}$) line.}
  \label{fig:Spec_RL_H42a}
\end{figure}

The last scenario is that the line emission at $\sim$ 85.69 GHz from the galactic nucleus is blended with other spectral lines.  The H42$\alpha$ line in the N region has a FWHM approximately twice the FWHM of the molecular lines in the same spectra window (e.g., $c$-C$_3$H$_2$, CH$_3$C$_2$H, and H$^{13}$CN).  While it is possible that the velocity dispersion of the ionized gas is higher than for the molecular gas in that region, it is also possible that the higher FWHM is indicative that the H42$\alpha$ line is blended with another line.   Figure \ref{fig:Spec_RL_H42a} shows that the $4_{32}$$\rightarrow$$4_{23}$ transition of $c$-C$_3$H$_2$  at 85.65 GHz is close to the frequency of the H42$\alpha$ line.  The much stronger $2_{12}$$\rightarrow$$1_{01}$ transition of $c$-C$_3$H$_2$ is detected at a rest frequency of 85.34 GHz (the left column of panels in Figure \ref{fig:Spec_RL}).   In the starburst/AGN composite galaxy, NGC~4945, \citet{Henkel2018} and \citet{Emig2020} have found that the H42$\alpha$ line emission is probably blended with the $c$-C$_3$H$_2$ ($4_{32}$$\rightarrow$$4_{23}$) line, and this contamination effect may increase the flux density by a factor of between $\sim$ 1.1 to 1.6 based on different assumptions.  If the H42$\alpha$ emission is blended with $c$-C$_3$H$_2$ emission in the nucleus of NGC~1808, the combined line emission could have been easier to detect than the H40$\alpha$ emission.  However, the two lines are so close that it is difficult to either confirm or quantify the relative fraction of the emission originating from the $c$-C$_3$H$_2$ in our spectra.  Additionally, it is unclear if the line ratios estimated from the NGC~4945 are applicable to NGC~1808.  If we exclude the channels covering the potential $c$-C$_3$H$_2$ ($4_{32}$$\rightarrow$$4_{23}$) line between 85.63 and 85.66~GHz, corresponding to a velocity range of $\sim$ 100 $\kms$, the H42$\alpha$ line flux will be reduced by $\sim$ 40 per cent, which is broadly consistent with the line flux ratios found in NGC~4945 \citep{Emig2020}.  Having said this, the millimeter continuum emission, which is dominated by free-free emission (see in Section \ref{sec:SED}), clearly peaks in the nucleus, which indicates the presence of the photoionized gas that would be expected to have recombination line emission associated with it.  This implies that it is more likely that the line emission at 85.69 GHz is predominantly H42$\alpha$ emission.

Given the discrepancies in the measurements of the H40$\alpha$ and H42$\alpha$ lines from the N region, we will treat quantities calculated from these lines as the lower and upper limits.  We also discuss how excluding the channels covering the potential $c$-C$_3$H$_2$ emission between 85.63 and 85.66~GHz may affect these quantities.

% Lines at rest-frame frequencies ranging from 85.68 to 85.70~GHz from splatalogue\footnote{\url{https://splatalogue.online/}}, Cyanoallene H$_2$CCCHCN $(17( 1,17)-16( 1,16))$, Ethylene Oxide $c-$H$_2$COCH$_2$ $13(10, 3)-13( 9, 4)$, Vinyl Cyanide CH$_2$CHCN (multi), Methyl Formate CH$_3$O$^{13}$CHO (TopModel), Methyl Mercaptan CH$_3$SH, vt$\leq$2
% \citet{Salak2016,Salak2017} presented CO (1-0), CO(3-2). \citet{Salak2018} presented HCN (1-0), H$_{13}$CN (1–0), HCO+ (1–0), H13CO+ (1–0), HOC+ (1–0), HCO+ (4–3), CS (2–1), C2H (1–0), with velocities of $\sim$ 990 km/s and FWHMs of $\sim$ 130 km/s measured within central 3'' were listed in Table 1.\citet{Salak2019} presented CO (1-0), CO (2-1), CO (3-2) with velocities of 996 km/s and FWHMs of $\sim$ 145 km/s and [C I] (1–0) with a velocity of 1000 km/s

\section{Spectral Energy Distribution} \label{sec:SED}

\begin{table}
\centering
\begin{minipage}{0.33\textwidth}
\caption{Continuum measurements for SED analysis.}\label{tab:SED_flux}
    \begin{tabular}{@{}cccc@{}}
    \hline
    \hline
      Rest frame &
      \multicolumn{3}{c}{Flux density} 
      \\
      frequency &
      T region & 
      N region & 
      S region \\
      (GHz) &
      (mJy) &
      (mJy) &
      (mJy) \\
    \hline
      1.52 & 182 $\pm$ 18 & 60 $\pm$ 6 & 29 $\pm$ 4\\
      4.91 & 99 $\pm$ 10 & 36 $\pm$ 4 & 16 $\pm$ 2\\
      8.50 & 46 $\pm$ 5 & 18 $\pm$ 2 & 8 $\pm$ 1\\
      86.58 & 26 $\pm$ 1 & 8 $\pm$ 0.4 & 5 $\pm$ 0.3\\
      87.72 & 28 $\pm$ 1 & 8 $\pm$ 0.4 & 6 $\pm$ 0.3\\
      98.55 & 25 $\pm$ 1 & 7 $\pm$ 0.4 & 5 $\pm$ 0.3\\
      100.79 & 25 $\pm$ 1 & 7 $\pm$ 0.4 & 5 $\pm$ 0.3\\
      102.24 & 24 $\pm$ 1 & 7 $\pm$ 0.4 & 5 $\pm$ 0.2\\
      114.24 & 22 $\pm$ 1 & 7 $\pm$ 0.3 & 5 $\pm$ 0.3\\
      219.96 & 66 $\pm$ 7 & 19 $\pm$ 2 & 14 $\pm$ 1\\
      229.86 & 69 $\pm$ 7 & 18 $\pm$ 2 & 16 $\pm$ 2\\
      233.90 & 71 $\pm$ 7 & 19 $\pm$ 2 & 16 $\pm$ 2\\
      245.47 & 80 $\pm$ 8 & 22 $\pm$ 2 & 18 $\pm$ 2\\
      344.95 & 202 $\pm$ 20 & 59 $\pm$ 6 & 49 $\pm$ 5\\
      357.41 & 249 $\pm$ 25 & 69 $\pm$ 7 & 63 $\pm$ 6\\
    \hline
    \end{tabular}
\end{minipage}
\end{table}

For our analysis, we need to identify the fraction of continuum emission in ALMA Band 3 that originates from free-free emission.  To do this, we use all the ALMA and VLA continuum data between 1 and 360 GHz in the T, N and S regions to construct SEDs, which we then fit with functions representing synchrotron, free-free, and thermal dust emission.  The continuum measurements used in our SED analysis are listed in Table \ref{tab:SED_flux}.  We model the synchrotron emission and thermal dust emission as power laws where the scales and spectral indices are set as free parameters.  We use the Gaunt factor ($g_{ff}$) given by \citet{Draine2011} to set the shape of the free–free emission.  This function is given as
\begin{equation}
  g_{ff} = 0.5535 \ln \left| 
  \left[ \frac{T_e}{\mbox{K}} \right]^{1.5}
  \left[ \frac{\nu}{\mbox{GHz}} \right]^{-1}
  Z^{-1}
  \right| -1.682.  
\label{eq:gff}
\end{equation} 
where $\nu$ is the frequency and $Z$ is the charge of the ions, which is set to 1.  The scale of the free-free emission was treated as a free parameter.  For fitting the SED, we set $T_e$ to 5000~K, which is comparable to the numbers derived in Section~\ref{sec:ET} as well as to electron temperatures derived in other nearby starburst galaxies \citep{Bendo2015, Bendo2016}; our measurements of the flux density from free-free emission varies by $\leq 5$ per cent when $T_e$ varies from 1000 to 10000~K.  
% We used a Monte Carlo approach to estimate the uncertainties of fitting parameters and individual percents of total emission. 

Using the {\sc lmfit} package (\citealt{Newville2022}), we first used a Levenberg-Marquardt algorithm to fit the SED in a logarithmic space, and then we applied the Maximum likelihood via Monte-Carlo Markov Chain method in 1,000 iterations to sample the posterior distributions of the fitting parameters.  Based on the fitting parameters, we calculated the fractions of synchrotron, free-free, and thermal dust emission to the total emission.  Our measurements and associated uncertainties are derived from the medians and half the difference between the 15.8 and 84.2 percentiles of the posterior distributions, which is equivalent to the inner 68\% of the distributions. 
% lmfit: the Non-Linear Least-Squares Minimization and Curve-Fitting for Python

\begin{figure*}
  \centering
  \includegraphics[width=0.33\textwidth]{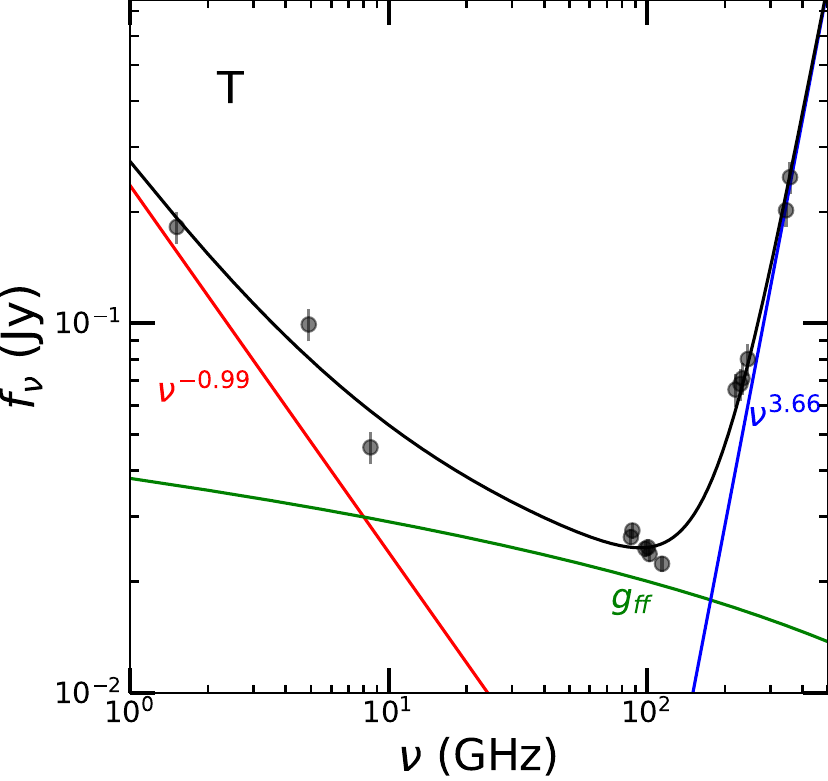} 
  \includegraphics[width=0.33\textwidth]{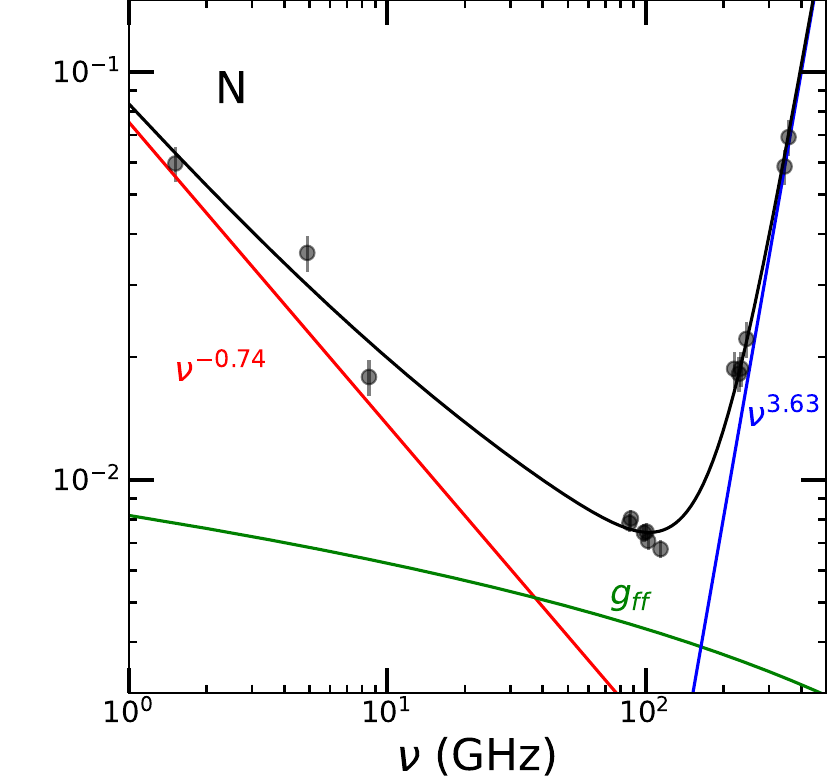} 
  \includegraphics[width=0.33\textwidth]{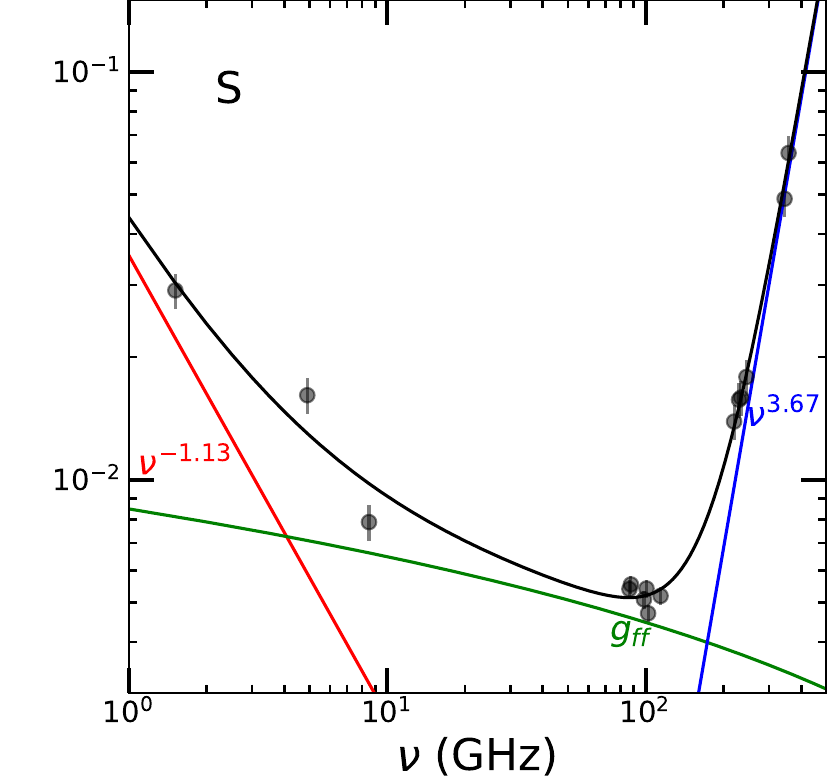} 
  \caption{The SEDs for the T, N and S regions of NGC~1808.  The black lines show the best fitting SED model for the data.  The red lines represent synchrotron emission.  The green lines represent free–free emission based on the Gaunt factor with $T_e$ fixed at 5000~K.  The blue lines correspond to thermal dust emission.}
  \label{fig:SED}
\end{figure*}

\begin{table*}
\centering
\begin{minipage}{0.975\textwidth}
\caption{SED analysis results.}\label{tab:SED_result}
    \begin{tabular}{@{}ccccccccc@{}}
    \hline
    \hline
       &
      \multicolumn{3}{c}{Synchrotron emission} &
      \multicolumn{2}{c}{Free-free emission} &
      \multicolumn{3}{c}{Thermal dust emission} 
      \\
      Region &
      Spectra & 
      Fraction at & 
      Fraction at &
      Fraction at & 
      Fraction at &
      Spectra & 
      Fraction at & 
      Fraction at \\
       &
      index &
      85.69~GHz (\%) &
      99.02~GHz (\%) &
      85.69~GHz (\%) &
      99.02~GHz (\%) & 
      index &
      85.69~GHz (\%) &
      99.02~GHz (\%) \\
    \hline
      T & $-0.99 \pm 0.16$ & $12 \pm 7$ & $10 \pm 7$
        & $83 \pm 8$ & $81 \pm 7$
        & $3.66 \pm 0.18$ & $5 \pm 1$ & $9 \pm 2$ \\
      N & $-0.74 \pm 0.14$ & $33 \pm 15$ & $30 \pm 14$ 
        & $62 \pm 15$ & $62 \pm 14$ 
        & $3.63 \pm 0.16$ & $5 \pm 1$ & $8 \pm 1$ \\
      S & $-1.13 \pm 0.20$ & $5 \pm 4 $ & $4 \pm 3 $ 
        & $89 \pm 4 $ & $86 \pm 4 $ 
        & $3.67 \pm 0.15$ & $6 \pm 1$ & $10 \pm 2 $ \\
    \hline
    \end{tabular}
\end{minipage}
\end{table*}

The best fitting parameters from the SED fits are listed in Table \ref{tab:SED_result}, and the SEDs and the best-fitting functions are shown in Figure \ref{fig:SED}.  According to the best fitting SED model for the data, the free-free emission dominates between 20 and 150~GHz and produces $\sim$80 per cent of the total observed emission (with an uncertainty of $\lesssim$8 per cent) at the rest frequencies of the H40$\alpha$ and H42$\alpha$ lines, while the synchrotron and dust emission produce the remaining $\sim$20 per cent of the emission.  The fraction of the total emission from synchrotron emission is almost seven times greater in the N region than in the S region, which may arise from the presence of the AGN or more supernovae in the galactic nucleus.  However, the widths of hydrogen recombination line are $\lesssim 300 \kms$, so even if an AGN is present in the nucleus, it should be a relatively weak energy source.

The spectral indices of the millimetre and radio continuum emission in the centre of NGC~1808 have been explored in several previous studies.  \citet{Dahlem1990} reported a mean spectral index of $-0.88$ for the nonthermal radiation in the central starburst based on 20 cm (1.5 GHz) and 6 cm (4.9 GHz) VLA data with a resolution of 20 arcsec.  The spectral index of $-0.99\pm0.16$ that we measured in the T region from our SED analysis is in good agreement with their measurement despite the fact that the data and methodology that we used to estimate the spectral index are different from theirs.  This agreement suggests that the synchrotron radiation dominates the 1.5~GHz emission from the central starburst.   
% It probably arises from the ageing effect of synchrotron radiation that causes spectra to steepen at high frequencies \citep[e.g.,][]{Davies2006}.  If we only use our 1.5 and 4.9 GHz data, we will get a special slope of $-0.51\pm0.13$, which is about 3$\sigma$ flatter than their value.  This difference is possibly because we use different data with smaller beam sizes.    

\citet{Salak2016} calculated spectra indices at each pixel from 102 and 114~GHz ALMA data and found that the indices varied significantly within the central region, ranging from $-3$ to 1 in the central 2 arcsec diameter region and equalling about $-1$ in the galactic centre.  However, these spectra index variations were affected by pixel-to-pixel variations in low signal-to-noise data.  \citet{Salak2017} performed a follow-up analysis of the 21 to 350 GHz SED of the central 2 arcsec diameter region using both ALMA and eVLA data, and they determined that 72 per cent of the 93~GHz emission originates from free-free emission, with 22 per cent arising from synchrotron and 6 per cent from dust.  The proportions of free-free emission, synchrotron emission and dust emission we measured in the nuclear region are comparable to those presented by \citet{Salak2017}, although the relative fractions of synchrotron emission and free-free emission measured by \citet{Salak2017} are about 10 per cent different from our measurements.  The variations between these results and ours are probably related to the different spatial resolution of data, the different apertures used to make the measurements, and the different methods used to fit SEDs (e.g., the Gaunt factor for free-free emission and unrestricted spectra indices of synchrotron and dust emission).  If we fixed the spectral indices of synchrotron emission and dust emission to the values used by \citet{Salak2017}, then we obtained 76 per cent of the 93~GHz emission originating from free-free emission and 19 per cent originating from synchrotron, which are consistent with their measurements.

More recently, \citet{Audibert2021} measured the spectral index from 8 to 93 GHz in the central 2 arcsec diameter region to be $-1$, which they interpreted as consistent with emission originating entirely from synchrotron emission in these bands, and they measured the spectral index from 350 to 500 GHz in the same region to be 3.6, which they used as evidence to indicate that dust emission dominated these bands.  Although we also conclude that the emission from 350 to 500~GHz is dominated by dust emission, our results show that the SED bends between 8 and 93 GHz (see Figure \ref{fig:SED}), which would indicate that using a single spectral index to characterize the emission between these bands, as done by \citet{Audibert2021}, will provide an inadequate description of the shape of the SED.

Aside from NGC~1808, the extragalactic SEDs from far-infrared continuum to radio continuum emission have only been published for M82 \citep{Condon1992,Peel2011}, NGC~253 \citep{Peel2011,Bendo2015}, NGC~4945 \citep{Peel2011,Bendo2016} and NGC~3256 \citep{Michiyama2020}.  
% \citet{Condon1992} analysed the SED from 1 to 6000~GHz for M82 using data from Klein 1998 and and Carlstrom 1991, and the SED analysis was updated by \citet{Peel2011}, who also analysed SED for two other nearby galaxies, NGC~253 and NGC~4945, using Planck and Wilkinson Microwave Anisotropy Probe data.  
% \citet{Bendo2015,Bendo16} continued on SED analysis to analyse the SED for NGC~253 and NGC~4959 using ALMA and VLA data.  
% \citet{Michiyama2020} analysed the SEDs from 5 to 224~GHz for the northern nucleus (2.5 arcsec) and southern nuclues (3.5 arcsec) of the dusty mergering galaxy, NGC~3256.
Although the proportions measured varied among the galaxies and although different groups obtained slightly different results for specific galaxies, these SED results still demonstrate that the dominant source of continuum emission from 84 to 116~GHz, corresponding to the frequency range of ALMA Band 3, is free-free emission.  Our SED results for NGC~1808 are largely consistent with this.  Additionally, these results from these other galaxies and our results for NGC 1808 show that dust emission is consistently the dominant source of emission seen in ALMA Bands 6 and 7.

\section{Electron Temperatures} \label{sec:ET}

The ratio of the hydrogen recombination line emission integrated over velocity $v$, $\int f_\nu$(line)$dv$, to the free–free emission, $f_\nu$(ff), can be written as 
\begin{equation}
\begin{split}
  \frac{\int f_\nu(\mbox{line}) dv}{f_\nu(\mbox{ff})} 
  \left[\frac{\mbox{Jy}}{\mbox{Jy km s}^{-1}}\right] = 
  \ \ \ \ \ \ \ \ \ \ \ \ \ \ \ \ \ \ \ \ \ \ \ \ \ \ \ \ \ \ \ \ \ \ \ \ \ \ \ \ \ \ \ \ \ \ \ \ \\
  4.38\times10^{33} g_{ff}^{-1}
  \left[\frac{\epsilon_\nu}{\mbox{erg s}^{-1}\mbox{ cm}^{-3}}\right]
  \left[\frac{\nu}{\mbox{~GHz}}\right]^{-1}
  \left[\frac{T_e}{\mbox{~K}}\right]^{0.5},
\label{eq:Te}
\end{split}
\end{equation}
which we can use to derive electron temperatures ($T_e$), as presented by \citet{Bendo2016} based on equations from \citet{Draine2011} and \citet{Scoville2013}.  $\epsilon_\nu$ is the emissivity from \citet{Storey1995}, and while it does not change significantly as a function of electron density ($n_e$) between $10^2$ and $10^5$ cm$^{-3}$, it varies strongly as a function of $T_e$ in the range of 500-30000~K.  We therefore use $\epsilon_\nu$ corresponding to case B recombination with a fixed $n_e$ value of $10^3$ cm$^{-3}$ but interpolate $\epsilon_\nu$ for different $T_e$.  In addition, $g_{ff}$ is also allowed to vary with $T_e$ for our analysis, although the maximum variation of $g_{ff}$ with $T_e$ is only $\sim 2 \times$ over the range of 2000-20000~K.  Two separate electron temperature measurements are made based on the two separate spectral lines.

% H40 alpha: 
% T: 1214.76064003  (1170.4082148   1126.05578958  1081.70336436  1037.35093914   992.99851392   948.6460887    904.29366348   859.94123825)   815.58881303   771.23638781   726.88396259
% N: 1214.76064003  (1170.4082148   1126.05578958  1081.70336436  1037.35093914  992.99851392   948.6460887)    904.29366348
% S: 992.99851392   (948.6460887    904.29366348   859.94123825   815.58881303)   771.23638781
% H42 alpha:
% T:  728.87609597   780.13050864   (831.38492132   882.639334  933.89374668   985.14815936  1036.40257204  1087.65698472  1138.91139739)  1190.16581007
% N:  882.639334  (933.89374668   985.14815936  1036.40257204  1087.65698472  1138.91139739)  1190.16581007
% S:  780.13050864   (831.38492132   882.639334  933.89374668)   985.14815936

\begin{table*}
\centering
\begin{minipage}{1\textwidth}
\caption{$T_e$ analysis results.}\label{tab:Te_result}
    \begin{tabular}{@{}ccccccccc@{}}
    \hline
    \hline
      Region &
      85.69~GHz& 
      H42$\alpha$ &
      H42$\alpha$ & 
      $T_e$ &
      99.02~GHz & 
      H40$\alpha$ &
      H40$\alpha$ & 
      $T_e$ \\
       &
      free–free & 
      line &
      line/free-free & 
      from H42$\alpha$ &
      free–free & 
      line &
      line/free-free & 
      from H40$\alpha$ \\
       &
      flux density &
      flux &
      emission ratio &
      line/free-free &
      flux density &
      flux &
      emission ratio &
      line/free-free \\
       &
      (mJy) &
      (mJy $\kms$) &
      (mJy $\kms$ mJy$^{-1}$) &
      (K) &
      (mJy) &
      (mJy $\kms$) &
      (mJy $\kms$ mJy$^{-1}$) &
      (K) \\
    \hline
      T & 
      $26 \pm 2$ & $1610 \pm 170$ & $63 \pm 9$ & $3800 \pm 500$ &
      $23\pm 2$ &  $980^\star\pm 200$ & $42 \pm 9$ & $6400^\star\pm 1500$ \\
      N & 
      $5\pm 1$ & $420 \pm 30$ & $81 \pm 20$ & $3000 \pm 700$ & 
      $5\pm 1$ & $90^\star\pm 50$ & $18 \pm 11$ & $13700^\star\pm 8000$ \\
      S &
      $5\pm 0.2$ & $200 \pm 30$ & $40 \pm 6$ & $5700 \pm 900$ & 
      $5\pm 0.2$ & $270 \pm 40$ & $57 \pm 9$ & $4900 \pm 700$ \\
    \hline
    \end{tabular}
    $^\star$ These values are affected by the non-detected H40$\alpha$ line emission in the N region.\\
    Note. These line flux and free-free flux density measurements do not include calibration uncertainties, as the ratios and the $T_e$ derived from the ratios should be independent of the calibration.  However, the uncertainties in the fraction of continuum emission that originates from free-free emission are incorporated into these data.
\end{minipage}
\end{table*}

Table \ref{tab:Te_result} lists all the free-free flux densities, spectra line fluxes and electron temperatures measured in the three regions shown in Figure~\ref{fig:Image_Aper}.  The spectra line fluxes are directly measured from the data (and not based on Gaussian fits).  To integrate the flux for the H42$\alpha$ lines in each spectra and for the H40$\alpha$ line in the S spectrum, we selected channels that would give line fluxes with optimized signal-to-noise ratios; the relativistic velocity widths covered by the channels in the T, N, and S spectra are 350 $\kms$, 250 $\kms$ and 150 $\kms$.  For the H40$\alpha$ lines in the T and N spectra, which have a low signal-to-noise, we used the same velocity ranges as used for the H42$\alpha$ line.  The free-free flux densities and spectra line fluxes within the N and S regions are approximately 40 per cent of that within the T region, which implies that the central 1~kpc diameter region also encompasses additional fainter star-forming regions. 

The $T_e$ measured in the T region is 6400~K based on the H40$\alpha$/99.02~GHz free-free emission ratio but only 3800~K based on the H42$\alpha$/86.59~GHz free-free emission ratio.  
The difference of $T_e$ arises from the missing H40$\alpha$ line emission in the N region, as discussed in Section \ref{subsec:spectra}.  If we remove the channels covering the $c$-C$_3$H$_2$ ($4_{32}$$\rightarrow$$4_{23}$) line (using 100 $\kms$ for the assumed line width), the electron temperatures from the H42$\alpha$/86.59~GHz free-free continuum ratio will increase to $5900\pm 1000$~K and $4700\pm 1100$~K for the T and N regions, respectively.  Since the $T_e$ values measured from the N region are not robust, we will not rely on them for the rest of our analysis.  We therefore use the average electron temperature of $5200\pm 500$~K from the T and S regions in the calculation of SFRs in Section~\ref{sec:SFR}.

The average electron temperature for the central starburst of NGC~1808 is comparable to that of the central region of the Milky Way \citep[e.g.,][]{Shaver1983,Paladini2004}, the central starburst of NGC~4945 \citep{Bendo2016} and the northern starburst nucleus of NGC~3256 \citep{Michiyama2020}, while it is slightly higher than that of the centre of NGC~253 \citep{Bendo2015}.  If we assume only half of the 86.59~GHz continuum emission originated from free-free emission, then the resulting electron temperature from the T region would be $\sim 2300$ K, which seems unreasonably low compared to that of nearby galaxies.  We can therefore infer that the free-free emission comprises more than half of the total 85.69~GHz continuum emission for the central starburst of NGC 1808.  
% The line/free-free emission ratio measured in the S region is slightly lower than that of NGC~253 and NGC~4945, while the ratio measured in the N region is significantly higher.  
% $T_e$ measured in the S region is $1800$~K higher than that measured in the N region based on 86.59~GHz emission.  These measurements are consistent with the starburst nucleus of NGC~3256 \citep{Michiyama2020}.   
% If we assume that the 86.59~GHz continuum emission was purely from free–free emission, then the $T_e$ is only 4300 K in the N region, which is still $1100$~K lower than the $T_e$ measured in the S region.  This increase of $T_e$ from the galactic centre to the circumnuclear star-forming region is in contrast to the trend of NGC~253 \citep{Bendo2015} and NGC~4945 \citep{Bendo2016}, but it is consistent with the positive correlation of $T_e$ with Galactocentric radius found in Milky Way \citep[e.g.,][]{Shaver1983,Paladini2004}.  
% The comparable $T_e$ between the S region and Galactic HII regions \citep[e.g.,][]{Shaver1983,Paladini2004} might also indicate that the circumnuclear region are dominated by photoionized gas.

\section{Star Formation Rates} \label{sec:SFR}

\subsection{SFR measurements from ALMA data} \label{subsec:SFR_ALMA}

The SFR can be calculated from the photoionizing photon production rate ($Q$) using
\begin{equation}
  \frac{\mbox{SFR}}{\mbox{M}_\odot~\mbox{yr}^{-1}}
   =7.29\times10^{-54}\frac{Q}{\mbox{s}^{-1}},
\label{eq:QtoSFR}
\end{equation}
where the conversion was derived from \citet{Murphy2011} using the {\sc Starburst99} model \citep{Leitherer1999} with solar metallicity ($Z_\odot$ = 0.020), a \citet{Kroupa2002} initial mass function (IMF) for a mass range of $0.1-100$~M$_\odot$ and no stellar rotation.  While different conversions that account for the stellar rotation from \citet{Leitherer2014} were used in previous studies \citep{Bendo2015,Bendo2016,Bendo2017}, these conversions are not widely used and also produce significantly different SFRs from other conversion equations.  To simplify our comparison of SFRs, we use a standard conversion that includes no rotation effects.
% [Guangwen to add some text on conversions using stellar rotation from Bendo et al. 2015, 2016, 2017, Leitherer 2011]

The hydrogen recombination line emission integrated over velocity and free–free emission can be related to the values to $Q$ by applying
\begin{equation}
  Q=\alpha_B n_e n_p V
\label{e_emtoq}
\end{equation}
from \citet{Scoville2013}.  $\alpha_B$ is the effective recombination coefficient from \citet{Storey1995}, which varies strongly with $T_e$, but changes little for $n_e$ values ranging from $10^2$ to $10^5$~cm$^{-3}$. % by a factor of $\sim$4 between 3000 and 15000~K
We therefore keep the electron density $n_e$=$10^3$~cm$^{-3}$ constant and use an $\alpha_B$ corresponding to the $T_e$ value of 5200~K, which is the average derived from the T and S region in Section~\ref{sec:ET}.  Following this, we can convert the free-free and recombination line emission to SFRs using
\begin{equation}
\begin{split}
\frac{\mbox{SFR}(\mbox{ff})}{\mbox{M}_\odot\mbox{ yr}^{-1}}=
  1.28\times10^{11}
  \ \ \ \ \ \ \ \ \ \ \ \ \ \ \ \ \ \ \ \ \ \ \ \ \ \ \ \ \ \ \ \ \ \ \ \ \ \ \ \ \ \ \ \ \ \ \ \ \ \ \ \ \\
  \times g_{ff}^{-1}
  \left[\frac{\alpha_B}{\mbox{ cm}^3\mbox{ s}^{-1}}\right]
  \left[\frac{T_e}{\mbox{~K}}\right]^{0.5}
  \left[\frac{D}{\mbox{ Mpc}}\right]^{2}
  \left[\frac{f_\nu(\mbox{ff})}{\mbox{Jy}}\right]
\end{split}
\label{eq:SFR_freefree}
\end{equation}
and
\begin{equation}
\begin{split}
  \frac{\mbox{SFR}(\mbox{line})}{\mbox{M}_\odot\mbox{ yr}^{-1}}=
  2.91\times10^{-23}
  \left[\frac{\alpha_B}{\mbox{ cm}^3\mbox{ s}^{-1}}\right]
  \ \ \ \ \ \ \ \ \ \ \ \ \ \ \ \ \ \ \ \ \ \ \ \ \ \ \ \ \ \ \\
  \times
  \left[\frac{\epsilon_\nu}{\mbox{erg s}^{-1}\mbox{ cm}^{-3}}\right]^{-1}
  \left[\frac{\nu}{\mbox{~GHz}}\right]
  \left[\frac{D}{\mbox{ Mpc}}\right]^{2}
  \left[\frac{\int f_\nu(\mbox{line}) dv}{\mbox{Jy km s}^{-1}}\right],
\end{split}
\label{eq:SFR_RL}
\end{equation}
where $D$ is the distance, and $f_\nu$(ff) and $\int f_\nu$(line)$dv$ are listed Table~\ref{tab:Te_result}. % The SFR(ff) is dependent upon the fraction of continuum emission from free-free emission determined in Section~\ref{tab:SED_result}, so it would increase by $\sim 24$ per cent if all the continuum emission was originated from free-free emission, and it would decrease by $\sim 38$ per cent if only half the continuum emission originated from free-free emission, which could cause extremely low $T_e$ (Section \ref{sec:ET}). 
Because the two different hydrogen line measurements from the galactic nucleus yield different results, SFR(line) is calculated only for comparison to SFR(ff).  
%In fact, SFR(line) and SFR(ff) are mathematically coupled together because the line-to-continuum ratio are coupled together with the $T_e$ measurements (Equation \ref{eq:Te}). 

\begin{table}
\centering
\begin{minipage}{0.44\textwidth}
\caption{SFR measurments from ALMA data.}\label{tab:SFR_ALMA}
    \begin{tabular}{@{}ccccc@{}}
    \hline
    \hline
    Region &
    85.69~GHz &
    H42$\alpha$ &
    99.02~GHz & 
    H40$\alpha$ \\ 
    & 
    free-free & 
    line &
    free-free &
    line \\
    & 
    (M$_\odot$ yr$^{-1}$) &
    (M$_\odot$ yr$^{-1}$) & 
    (M$_\odot$ yr$^{-1}$) &
    (M$_\odot$ yr$^{-1}$) \\
    \hline
    T &
      $3.2 \pm 0.4$ & % $3.17 \pm 0.39$ & 
      $4.5 \pm 0.7$ & % $4.49 \pm 0.72$ \\
      $3.0 \pm 0.4$ & % $2.97 \pm 0.35$ &
      $2.4^a \pm 0.6$ \\ % $2.35 \pm 0.56$ &
    N &
      $0.63 \pm 0.16$ &
      $1.16 \pm 0.15$ &
      $0.63 \pm 0.15$ &
      $<0.65^b$ \\% $0.22 \pm 0.13$ $^*$
    S &
      $0.62 \pm 0.05$ &
      $0.55 \pm 0.10$ &
      $0.59 \pm 0.05$ &
      $0.64 \pm 0.11$ \\
    \hline
    \end{tabular}
    $^a$ This value is affected by the non-detected H40$\alpha$ line emission. \\
    $^b$ This upper limit is estimated with 5$\sigma$ uncertainty.
\end{minipage}
\end{table}

The final SFRs measured from the ALMA data are listed in Table~\ref{tab:SFR_ALMA}.  The average of the free-free SFRs integrated over the central starburst (T region) is $3.1 \pm 0.3$~M$_\odot$ yr$^{-1}$; we use this average later in comparisons to SFRs from other bands.  While the area of the combined N and S regions is only 15 per cent of the T region, the emission from these two regions contributes approximately 40 per cent of the total SFR in the central starburst.  
% In addition, if we consider the stellar rotation, the SFR decreases by an additional 26\% (Equation \ref{eq:QtoSFR}).

The total SFR within the central starburst is broadly consistent with the recent published SFR of $2.5-3.5$~M$_\odot$ yr$^{-1}$ within the central 30 arcsec region using the ALMA 93~GHz emission, which they treat as dominated by free–free emission \citep{Salak2017}; the SFR values have been rescaled for a distance of 9.5~Mpc for comparison to our measurements.  In addition, \citet{Salak2017} also measured an SFR of $0.16-0.22$~M$_\odot$ yr$^{-1}$ in a 2 arcsec aperture.  Given the difference in aperture size, the nuclear SFR of $0.63$~M$_\odot$ yr$^{-1}$ that we obtained in an 6~arcsec aperture is broadly comparable to their measurement.  
% Salak2017 for the central 2" region, 0.28 using 100% 93 GHz (Table 5), for r<15", 4.5 using 100% 93 GHz, D=10.8 Mpc 

After deriving the SFR from the millimetre data, we calculated SFRs from radio, infrared, H$\alpha$ and ultraviolet data using the same {\sc Starburst99} model with the \citet{Kroupa2002} IMF.  Since the beam FWHMs in the infrared data are significantly larger than the aperture sizes of the N and S regions, we only compared the SFRs measured from multi wavebands within the central 1~kpc starburst (T region).  

\begin{table}
\centering
\begin{minipage}{0.475\textwidth}
\caption{SFR measurements for the central 1~kpc in NGC~1808.}\label{tab:SFR}
    \begin{tabular}{@{}lcc@{}}
    \hline
    SFR Metrics &
    Flux or flux density &
    SFR (M$_\odot$ yr$^{-1}$) \\
    \hline
    85.69~GHz free-free & $26 \pm 2$ mJy & $3.2 \pm 0.4$  \\
    H42$\alpha$ line & $1.6\pm 0.2$ Jy$\kms$ & $4.5 \pm 0.7$  \\
    99.02~GHz free-free & $23 \pm 2$ mJy & $3.0 \pm 0.4$  \\
    H40$\alpha$ line & $1.0^\star\pm 0.2$ Jy$\kms$ & $2.3^\star \pm 0.6$  \\
    1.5~GHz continuum & $344\pm34$ mJy & $2.3 \pm 0.2$ \\ 
    TIR continuum & $8.7\pm 0.5\times 10^{14}$ Jy Hz & $3.6 \pm 0.2$ \\
    160 $\mu$m infrared continuum & $86 \pm 7$ Jy & $2.4 \pm 0.2$  \\
    70 $\mu$m infrared continuum & $122 \pm 7$ Jy & $3.3 \pm 0.2$  \\
    22 $\mu$m infrared continuum & $11.9 \pm 0.7$ Jy & $3.6 \pm 0.2$ \\
    $\ha$ (corrected using TIR) & $2.2\pm 0.1\times 10^{12}$ Jy Hz & $1.3\pm 0.1$ \\
    $\ha$ (corrected using 22 $\mu$m) & $3.4\pm 0.2\times 10^{12}$ Jy Hz & $2.0 \pm 0.1$ \\
    NUV (corrected using TIR) & $2.4\pm 0.1\times 10^{14}$ Jy Hz & $1.7 \pm 0.1$ \\
    NUV (corrected using 22 $\mu$m) & $3.7\pm 0.2\times 10^{14}$ Jy Hz & $2.7 \pm 0.2$ \\
    FUV (corrected using TIR) & $4.0\pm 0.2\times 10^{14}$ Jy Hz & $1.9 \pm 0.1$ \\
    FUV (corrected using 22 $\mu$m) & $6.3\pm 0.4\times 10^{14}$ Jy Hz & $3.0 \pm 0.2$ \\
    \hline
    \end{tabular}
    $^\star$ These values are affected by the non-detected H40$\alpha$ line emission.\\
\end{minipage}
\end{table}

\begin{figure*}
  \centering
  \includegraphics[width=1\textwidth]{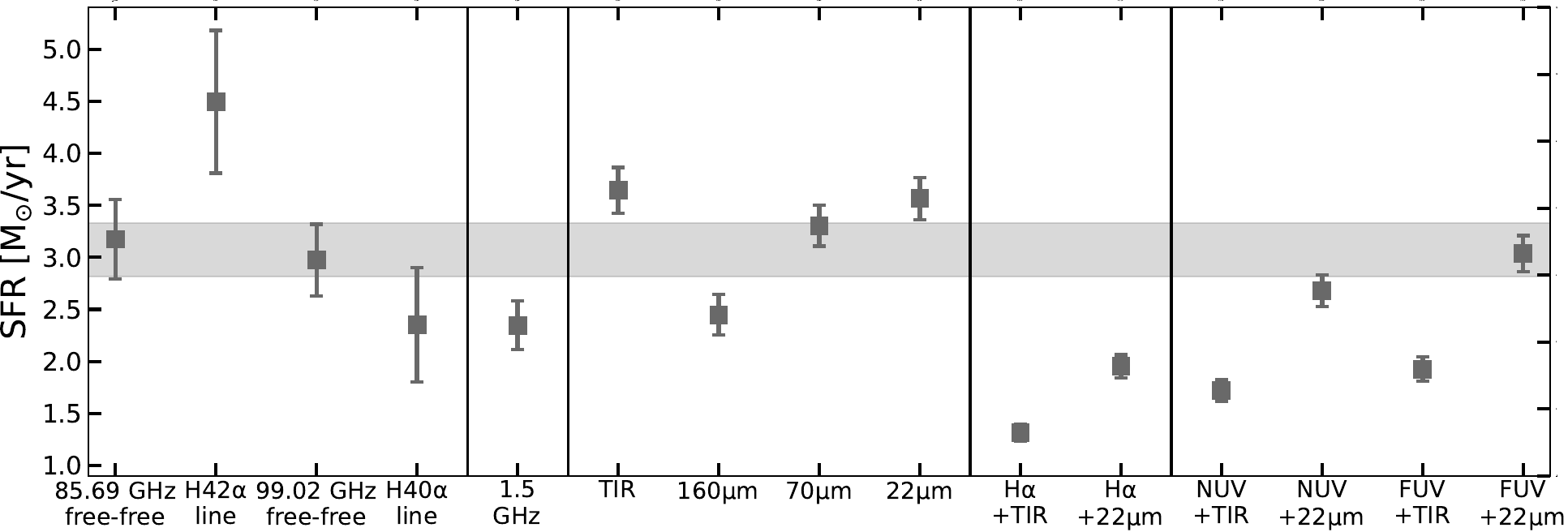} 
  \caption{SFR measurements for the central starburst in NGC~1808 (T region).  The grey shaded region corresponds to the mean within 1$\sigma$ uncertainties of SFRs from the H42$\alpha$ line and 85.69~GHz free-free emission.}
  \label{fig:SFR}
\end{figure*}

All of the SFR measurements for the central starburst in NGC~1808 are summarized in Table \ref{tab:SFR} and illustrated in Figure \ref{fig:SFR}.  In the following sections, we will discuss the comparisons of the SFR measurements from ALMA and the SFR measurements from radio, infrared, $\ha$ and ultraviolet data specifically.

\subsection{Comparisons of ALMA and radio SFRs}

For calculating SFRs based on the radio flux density, we use the integrated measurement in the T aperture of $344 \pm 34$ mJy from the 1.5~GHz VLA data (Section \ref{subsec:VLA}) and 
\begin{equation}
  \frac{\mbox{SFR}(1.5\mbox{GHz})}{\mbox{M}_\odot\mbox{ yr}^{-1}}
  = 0.0760 \left[\frac{D}{\mbox{Mpc}}\right]^{2}
  \left[\frac{f_\nu(1.5 \mbox{ GHz})}{\mbox{Jy}}\right]
  % 6.35\times10^{-29} \frac{L_\nu(1.5\mbox{GHz})}{\mbox{erg}\mbox{ s}^{-1}\mbox{ Hz}^{-1}}
\label{eq:sfr_radio}
\end{equation}
given by \citet{Murphy2011}, and we obtained an SFR of $2.3 \pm 0.2$ M$_\odot$ yr$^{-1}$.%  Similarly, we can obtained SFRs of $0.53 \pm 0.05$ M$_\odot$ yr$^{-1}$ for the N region.
The SFR estimated from the 1.5~GHz emission is approximately 75 per cent of the SFR from the free-free emission; the difference between the two values is $\sim$2$\sigma$.  This difference was most likely caused by the diffusion through the interstellar medium of the cosmic rays producing the synchrotron radiation \citep[e.g.,][]{Murphy2006}, which would make the radio emission appear broad compared to other star formation tracers.  In addition, if a radio-emitting AGN is present in the galactic nucleus, the SFR from radio emission that we calculated could be expected to be much higher than what we derived from the free-free emission.  Since the SFR from the 1.5~GHz band is lower than what we get from free-free emission, this implies that the AGN is not a significant energy source in radio bands.
% As a star formation tracer, the 1.5~GHz emission is based on the empirical correlation between integrated infrared and radio emission \citep[e.g.,][]{Murphy2011}, which may not provide very accurate SFR measurement in this galaxy. 

\subsection{Comparisons of ALMA and infrared SFRs}
%\subsection{Comparisons of ALMA and mid-infrared, far-infrared and total infrared SFRs}

\subsubsection{Infrared photometry}

Unlike millimetre and radio data, infrared emission is seen in a compact central region as well as diffuse structures distributed throughout the optical disc (Figure \ref{fig:Image_other}).  As a consequence, we need to remove this extended emission by applying local background subtractions when performing the measurements.  In addition, we need to apply the aperture corrections to the infrared flux densities because the infrared data is not well resolved, even in the T region. 
% with the same background annulus used in the local background subtractions
% especially for SPIRE 250~$\mu$ data, a larger aperture is needed.

In the {\it WISE} 22~$\mu$m image, we followed the procedures used in the WAPPco photometry system to produce the {\it WISE} All-Sky Data Release Products.  We first measured the target flux density from a 22~arcsec aperture corresponding to the T region.  After this, we subtracted the median background within an annulus with radii of 50–70 arcsec and applied an aperture correction of 2.77 that, based on the {\it WISE} beam provided by \citet{Cutri2013}, is appropriate for this target and background aperture combination.  No colour correction is needed in {\it WISE} 22~$\mu$m data because this effect is negligible ($\ltsim$1 per cent) from unity for power law spectra ranging from $\nu^{-3}$ to $\nu^3$ \citep{Wright2010}.  Following these calculations, we obtained a 22~$\mu$m flux density of $11.9 \pm 0.7$ Jy.  %For comparison, if we performed measurements the standard aperture with a radius of 16.5 arcsec in the WAPPco photometry system with the corresponding aperture correction and the same local background annulus, the {\it WISE} 22~$\mu$m flux density would increase by about 5 per cent. 
% Test1: flux density of $12.53 \pm 0.73$ Jy use standard aperture after applying the corresponding aperture correction of 1.76 and same local background subtraction.
% Test2: flux density without local background subtraction, $13.99 \pm 0.82$ Jy
% our measurment falls within $\sim$3\% of the {\it WISE} 22~$\mu$m measurement.
 
In the PACS 70, 100 and 160~$\mu$m images, we first measured the target flux densities within the 22~arcsec apertures, subtracted the median background measured within annuli with radii of 61-70~arcsec and multiplied by aperture correction factors of 1.26, 1.32 and 1.52, which come from the PACS Ancillary Data Products \citep{Lutz2015}\footnote{The data products are available from \url{http://archives.esac.esa.int/hsa/legacy/ADP/PSF/PACS/PACS-P/EEF\_Data/} and the relevant document is available from \url{http://archives.esac.esa.int/hsa/legacy/ADP/PSF/PACS/PACS-P/bolopsf\_22.pdf} }.  Next, we multiplied our measurements by colour correction factors of 1.06, 1.00 and 1.03 (corresponding to the PACS 70, 100 and 160~$\mu$m bands, respectively), which were derived by \citet{Bendo2016} from values given by \citet{Muller2011} for modified blackbodies with emissivities scaling as either $\nu$ or $\nu^{2}$ and with temperatures ranging from 20 to 50~K.  Following this, we obtained flux densities of $122\pm7$, $128\pm8$ and $86\pm7$ Jy at 70, 100 and 160~$\mu$m, respectively. 
%For comparison, if we used a background annulus with radii of 20-30~arcsec to apply the local background subtraction, then the far-infrared flux densities wouldn't decrease by more than 5 per cent. 
% Test1: flux densities with local background subtraction based on radii of 20-30 arcsec, $120.79\pm7.25$, $124.37\pm7.46$ and $81.14\pm6.50$ Jy
% Test2: flux density without local background subtraction, 
 
In the SPIRE 250~$\mu$m image, we first measured the target flux densities from the aperture with a radius of 22~arcsec, and then subtracted the median background within an annulus with radii of 30-60~arcsec.  Following the SPIRE Data Reduction Guide \citep{Spire2016}, we multiplied this value by an aperture correction of 1.27.  Next, we multiplied the flux density by the colour correction factor of 0.955 for SPIRE 250~$\mu$m dust emission given by \citet{Bendo2016}, who derived this factor based on values from the SPIRE Handbook \citep{Valtchanov2018} for point-like modified blackbodies with emissivity indices of $\nu^{-1.5}$ or $\nu^{-2}$ and temperatures ranging from 15 to 40~K.  We obtained a final 250~$\mu$m flux density of $32.5 \pm 1.8$ Jy. 
%For comparison, if we applied no local background subtraction, the flux density would increase by 6 per cent.  
% Test1: flux densities within T aperture, 25.97 \pm 1.43 Jy, 20\% difference.
% Test2: flux density without local background subtraction, $34.47\pm1.90$ Jy.
% colour correction in (Equation 5.20) Section 5.6 & 5.7 of SPIRE HANDBOOK. 

\subsubsection{Comparison of ALMA and total infrared SFRs}

Using the infrared measurements from 22~$\mu$m to 500~$\mu$m and 
\begin{equation}
\begin{split}
  F(\mbox{TIR}) = 2.023\nu f_\nu(22\mu\mbox{m}) + 0.523\nu f_\nu(70\mu\mbox{m}) 
  \ \ \ \ \ \ \ \ \ \ \ \ \ \ \ \ \ \ \ \ \ \ \ \ \\
  + 0.390\nu f_\nu(100\mu\mbox{m}) + 0.577\nu f_\nu(160\mu\mbox{m}) + 0.721\nu f_\nu(250\mu\mbox{m}) 
\end{split}
\label{eq:F_TIR}
\end{equation}
given by \citet{Galametz2013}, we obtained a total infrared (TIR) flux of $8.7\pm 0.5\times 10^{14}$ Jy Hz, equivalent to a TIR luminosity of $2.5\pm0.2 \times 10^{10}\Lsun$.  Using the conversion  
\begin{equation}
  \frac{\mbox{SFR}(\mbox{TIR})}{\mbox{M}_\odot\mbox{ yr}^{-1}}=
  4.66\times10^{-17}\left[\frac{D}{\mbox{Mpc}}\right]^{2}
  \left[\frac{F(\mbox{TIR})}{\mbox{Jy Hz}}\right]
 % 3.88\times10^{-44} \frac{L(\mbox{TIR})}{\mbox{erg}\mbox{ s}^{-1}}
\label{eq:sfr_TIR}
\end{equation}
from \citet{Murphy2011}, we obtain an SFR of $3.6 \pm 0.2$ M$_\odot$ yr$^{-1}$. % Using Rieke 2009 calibration, only 1.4 Msun/yr

The TIR flux yields an SFR approximately 15 per cent (or 2 $\sigma$) higher than the ALMA free-free SFR.  Given the uncertainties of the infrared photometry as well as  the additional uncertainties in the TIR flux calculation \citep[typically 4 per cent but potentially higher;][]{Galametz2013}, the difference is not significant.  

The assumption behind using the TIR flux to calculate SFR is that the TIR flux probes the emission absorbed and re-radiated by dust from star-forming regions that produce most of the bolometric luminosity of galaxies.  Additionally, \citet{Kennicutt2012} indicate that the conversion between TIR flux and SFR depends on the relative fraction of the stellar emission that is absorbed by dust.  The conversion from \citet{Murphy2011} is based on dust absorbing all energy shortwards of the Balmer break at 3646~\AA but not at longer wavelength.  However, in very dusty regions like the centre of NGC~1808, a significant fraction of the starlight at longer wavelengths may also be absorbed, thus leading to overestimated SFRs from the TIR.

Other non-photoionizing stars, ranging from ultraviolet-bright stars with ages just over 10~Myr to much older evolved red stars, may also heat interstellar dust \citep[e.g.,][]{Hao2011}.  However, in the centre of NGC 1808, our results indicate that energy absorbed by dust from those stars would produce 15 per cent of the TIR emission.  Additionally, energy absorbed by interstellar dust from evolved stars would tend to be emitted at longer wavelengths \citep[e.g.,][]{Kennicutt1998}, but the SFR calculated from the 160~$\mu$m data is relatively low in comparison (see Section \ref{subsubsec:SFR_monIR}), which implies that dust heating by evolved stars is not significant.

An AGN could also potentially produce additional infrared emission.  However, if an AGN were the dominant infrared emission source in this galaxy, then the TIR SFR would be expected to be much higher than the ALMA SFR.  Since the discrepancy between TIR SFR and ALMA SFR is only 15 per cent, and the SFR from the TIR emission is 1.5 times higher than the SFR from the 1.5 GHz emission (which would also be expected to be high if an AGN were present), we can conclude that any AGN that is present produces a relatively small fraction of the total infrared emission from NGC~1808 and that star-forming regions are the main source of infrared emission from this galaxy.

\subsubsection{Comparisons of ALMA and monochromatic infrared SFRs}
\label{subsubsec:SFR_monIR}

Beside the conversion from TIR emission to SFR, several conversions from monochromatic infrared dust emission to SFR are also widely used. %, though two main assumptions that the shape of infrared SED keeps unchanged and the monochromatic infrared luminosities scale linearly with the $L(\mbox{TIR})$ need to be added.
The conversion from {\it WISE} 22~$\mu$m emission to SFR is the same as that from other mid-infrared data \citep[e.g.,][]{Lee2013}.  Using one of the most widely used relations, 
\begin{equation}
  \frac{\mbox{SFR}(22\mu\mbox{m})}{\mbox{M}_\odot\mbox{ yr}^{-1}}=
  2.44\times10^{-16} \left[\frac{D}{\mbox{Mpc}}\right]^{2}
  \left[\frac{\nu f_\nu(22\mu\mbox{m})}{\mbox{Jy Hz}}\right]
  % 2.04\times10^{-43} \frac{\nu L_\nu(22\mu\mbox{m})}{\mbox{erg}\mbox{ s}^{-1}},
\label{eq:sfr_22}
\end{equation}
given by \citet{Rieke2009}, we obtained an SFR of $3.6 \pm 0.2$ M$_\odot$ yr$^{-1}$.  In this case, the corresponding mid-infrared luminosity is  $4.6\pm0.3\times10^{9}\mbox{L}_\odot$, which falls within the range of $6\times10^{8}$ to $1.3\times10^{10}\mbox{L}_\odot$ over which this relation should be applicable.  % $4.56\pm0.26\times10^{9}\mbox{L}_\odot$

The 70~$\mu$m and 160~$\mu$m measurements can also be used to calculate SFRs.  Using 
\begin{equation}
  \frac{\mbox{SFR}(70\mu\mbox{m})}{\mbox{M}_\odot\mbox{ yr}^{-1}}=
  7.04\times10^{-17} \left[\frac{D}{\mbox{Mpc}}\right]^{2}
  \left[\frac{\nu f_\nu(70\mu\mbox{m})}{\mbox{Jy Hz}}\right]
 %5.88\times10^{-44} \frac{\nu L_\nu(70\mu\mbox{m})}{\mbox{erg}\mbox{ s}^{-1}}
\label{eq:sfr_70}
\end{equation}
and 
\begin{equation}
  \frac{\mbox{SFR}(160\mu\mbox{m})}{\mbox{M}_\odot\mbox{ yr}^{-1}}=
  1.71\times10^{-16} \left[\frac{D}{\mbox{Mpc}}\right]^{2}
  \left[\frac{\nu f_\nu(160\mu\mbox{m})}{\mbox{Jy Hz}}\right]
 % 1.43\times10^{-43} \frac{\nu L_\nu(160\mu\mbox{m})}{\mbox{erg}\mbox{ s}^{-1}}
\label{eq:sfr_160}
\end{equation}
from \citet{Calzetti2010}, we obtained an SFR$(70\mu\mbox{m})$ of $3.3 \pm 0.2$ M$_\odot$ yr$^{-1}$ and an SFR$(160\mu\mbox{m})$ of $2.4 \pm 0.2$ M$_\odot$ yr$^{-1}$ for the central 1~kpc region.  In this case, the corresponding 70~$\mu$m and 160~$\mu$m luminosities are of $1.5\pm0.1\times10^{10}\mbox{L}_\odot$ and $4.5\pm0.4\times10^{9}\mbox{L}_\odot$, both of which fall within the ranges where these relations should apply.

The 22~$\mu$m and 70~$\mu$m flux densities yield SFRs 20 per cent and 10 per cent higher than the ALMA SFR, while the 160~$\mu$m flux density yields an SFR 17 per cent lower than the ALMA SFR.  The conversions from monochromatic infrared emission to SFR are based on assumptions that the shape of infrared SED remains unchanged and that the monochromatic infrared emission scales linearly with the TIR emission (as well as the additional assumptions involved in converting from TIR emission to SFR).  Considering the calibration uncertainties of the conversions are 0.2 dex, 0.2 dex and 0.4 dex for 22 ~$\mu$m, 70~$\mu$m and 160~$\mu$m emission, respectively \citep[][]{Rieke2009,Calzetti2010}, the discrepancies among these SFRs or between them and the ALMA free-free SFR are not significant.  

However, the calculated SFR still decreases notably with wavelength.  This relation may arise because the dust within NGC~1808 is relatively hot compared to the star-forming and starburst galaxies used by \citet{Calzetti2010}. This trend has also been found in NGC~5253, which is a dwarf starburst galaxy with abnormally hot dust \citep{Bendo2017}.    % In addition, calibration issues could be another reason for these discrepancies between the ALMA SFR and the far-infrared SFR. 
% The SFRs from 160~$\mu$m bands can be inaccurate if the monochromatic infrared emission from these bands cannot scale linearly with TIR emission.  

\subsection{Comparisons of ALMA, PAH and Br$\gamma$ SFRs}
% \subsection{Comparisons of ALMA and previous published near- and mid-infrared SFRs}

Some researchers have proposed using polycyclic aromatic hydrocarbon (PAH) emission to measure the SFR \citep[e.g.,][]{Diamond-Stanic2010, Diamond-Stanic2012}, which would be advantageous for measuring SFRs in locations where AGN produce much of the bolometric luminosity.  In NGC~1808, \citet{Esquej2014} measured an SFR of $0.55 M_\odot$ yr$^{-1}$ (rescaled for a distance of 9.5 Mpc) from the 11.3 $\mu$m PAH emission in an aperture comparable to our T region.  This is $\sim$5$\times$ lower than the SFRs from our ALMA and infrared data.  PAH emission is generally expected to be suppressed relative to hot dust emission and other commonly used star formation tracers, most likely because the PAHs are more easily destroyed by the hard ultraviolet radiation fields in star-forming regions \citep[e.g.,][]{Helou2004,Calzetti2005,Calzetti2007,Bendo2006,Bendo2008,Povich2007,Lebouteiller2007,Prescott2007,Jones2015,Egorov2023}. Even though the PAH 11.3~$\mu$m feature specifically was expected to be a more robust star formation tracer than other PAH lines \citep{Diamond-Stanic2010}, it works poorly as an SFR metric in the centre of NGC~1808. 

\citet{Krabbe1994} and \citet{Kotilainen1996} used the Br$\gamma$ line at 2.16~$\mu$m and radio continuum emission to calculate the SFRs for the NGC~1808 (although their SFRs have to be rescaled to correspond to a distance of 9.5 Mpc for comparison to our measurements).  Our ALMA nuclear SFR of $0.63$~M$_\odot$ yr$^{-1}$ falls within the SFR range of 0.2-2.0~M$_\odot$ yr$^{-1}$ measured by \citet{Krabbe1994}; note that their range in values is based on applying different lower mass cutoffs for the IMF.  The ALMA nuclear SFR is also consistent with the value of 0.7~M$_\odot$ yr$^{-1}$ measured by \citet{Kotilainen1996} within a comparable aperture. More recently, \citet{Busch2017} measured an SFR of 0.11~M$_\odot$ yr$^{-1}$ using the Br$\gamma$ line within a 1.5" diameter aperture after rescaling for a distance of 9.5 Mpc. Considering the difference in aperture size, our ALMA nuclear SFR is comparable to their measurement.   
% In addition, the starburst is the dominant mechanism contributing to the nuclear mid-IR emission \citep{Esquej2014}. 

%[Helou 2003, Bendo 2006 2008, Calzetti 2005 2007, Prescott 2007]

% SFR (NED) 4.7 from IR  
% Krabbe (1994), 2" region, Spot1 0.8, Spot5 1.2 with different IMF, using same IMF may need x 0.32 using the dereddened Br$\gamma$ line emission, D = 10.9 Mpc
% Kotilainen (1996) 0.53 Msun/yr, 2" diameter for the nucleus, 2.2 for the 6" diameter using the dereddened Br$\gamma$ line emission, D = 16.4 Mpc 
% Esquej (2014), NGC 1808, circ 20-24", nucleus/circ ratio: 0.17, nucleus 0.35", 0.21 Msun/yr from 11um PAH emission at D = 14.3 Mpc
% Busch (2017), 0.2 Msun/yr derived within a smaller region of 1.5" diameter, 0.728 Msun/yr within the sum of all 11 regions within 8" arcsec region (Table 3), using the Br$\gamma$ line of hydrogen based on the calibration from Panuzzo 2003 (a Salpeter IMF with m_inf = 0.15 Msun and m_up = 120 Msun and solar metallicity), D = 12.8 Mpc   

\subsection{Comparisons of ALMA and $\ha$ SFRs}
\label{subsec:SFR_ha}

For calculating SFRs from the $\ha$ flux, we first used pPXF to perform measurements with Gaussian emission line templates to the continuum-subtracted spectrum integrated over the central 22~arcsec region (see more in Section \ref{subsec:ha}).  This gives us a $\ha$ flux of $1.3\pm 0.1 \times 10^{11} {\mbox{Jy Hz}}$ and a corresponding luminosity of $3.8\pm0.2\times10^{6}\mbox{L}_\odot$. 
% $(1 {\mbox{ Jy Hz}} = 10^{-23} {\mbox{ erg}\mbox{ s}^{-1}\mbox{cm}^{-2}} $)  
Next, we corrected the $\ha$ flux for the intrinsic dust extinction using
\begin{subequations}
\begin{align}
 & F({\mbox{\ha}})_{\mbox{corr}} = F({\mbox{\ha}})_{\mbox{obs}} + 0.0024 F({\mbox{TIR}}), \label{eq:ha_cor_TIR} \\
 & F({\mbox{\ha}})_{\mbox{corr}} = F({\mbox{\ha}})_{\mbox{obs}} + 0.020 \nu f_\nu(22\mu{\mbox{m}}) \label{eq:ha_cor_22}
\end{align}
\end{subequations}
given by \citet{Kennicutt2009} and obtained $\ha$ fluxes of $2.2\pm 0.1\times 10^{12}$ Jy Hz corrected using the TIR flux and $3.4\pm 0.2\times 10^{12}$ Jy Hz corrected using the mid-infrared flux.  Following this, we used 
\begin{equation}
  \frac{\mbox{SFR}(\mbox{\ha})}{\mbox{M}_\odot\mbox{ yr}^{-1}}=6.43\times10^{-15}
  \left[\frac{D}{\mbox{ Mpc}}\right]^{2}
  \left[\frac{F(\mbox{H}\alpha)}{\mbox{Jy Hz}}\right]
%  5.37\times10^{-42} \frac{L(\mbox{\ha})}{\mbox{erg}\mbox{ s}^{-1}}
\label{eq:sfr_Ha}
\end{equation}
from \citet{Murphy2011} to obtain SFRs of $1.3\pm 0.1$ ${\mbox{M}_\odot\mbox{ yr}^{-1}}$ and $2.0 \pm 0.1$ ${\mbox{M}_\odot\mbox{ yr}^{-1}}$ corrected using TIR and mid-infrared emission, respectively. 

The extinction-corrected $\ha$ using TIR and mid-infrared fluxes yield SFRs 2.4$\times$ and 1.6$\times$ lower than the ALMA free-free SFR.  Considering the uncertainties in the conversions of the combination of $\ha$ and infrared emission to SFRs are $\sim$ 0.3 dex \citep{Kennicutt2009}, these discrepancies are significant.  

The conversions from the combination of $\ha$ and infrared emission to SFRs fail in this situation probably because of the heavy dust-obscuration of the central starburst in NGC~1808.  The observed TIR-$\ha$ flux ratio is $\sim 6600$ in the central region, which exceeds the applicability range of 45-3150 for the star-forming galaxies used in the \citet{Kennicutt2009} sample.  In Equations \ref{eq:ha_cor_TIR} and \ref{eq:ha_cor_22}, the observed $\ha$ flux is only equivalent to 6 per cent and 4 per cent of the TIR and 22$\mu$m terms.  This implies that the conversion of the extinction-corrected $\ha$ flux (Equation \ref{eq:sfr_Ha}) is effectively an indirect conversion from infrared flux to SFR.  As a result, although combining $\ha$ with infrared emission to measure SFR is widely and effectively used for star-forming galaxies, it still needs to be used with caution when applied to dusty starburst galaxies like NGC~1808, which may be the types of objects more commonly seen in the high-redshift universe.  
% If we applied the correction method that should only be used for $\hii$ regions for this galaxy by changing the coefficient used in Equation \ref{eq:ha_cor_22} into $0.031\pm 0.006$ given by \citet{Kennicutt2009}, then we obtained an SFR of $2.98\pm 0.17 {\mbox{M}_\odot\mbox{ yr}^{-1}}$, which is consistent with the ALMA SFR.  This consistency implies the stellar populations in the central starburst of NGC~1808 is more likely to be typical $\hii$ regions rather than star-forming galaxies. 

Aside from the issues with the extinction-corrected $\ha$ not yielding accurate SFRs for NGC~1808, it is also worth pointing out that the SFR from the extinction-corrected $\ha$ using mid-infrared emission is $\sim$ 1.5 higher than that using TIR emission.  This discrepancy may arise from the dust within the central starburst within NGC~1808 being relatively hot compared to the star-forming galaxies, so the 22~$\mu$m flux density will be relatively high in comparison to the TIR flux (as is also discussed in Section~\ref{subsubsec:SFR_monIR}) and will lead to a larger correction to the $\ha$ flux.

\subsection{Comparisons of ALMA and ultraviolet SFRs}

In the GALEX FUV and NUV images, we first measured the target flux densities from a 22~arcsec aperture corresponding to the T region.  After this, we subtracted the median background within an annulus with radii of 190-260 arcsec and applied correction factors of 1.15 and 1.23 for the foreground extinction (see more in Section \ref{subsec:UV}). % Extinction: FUV, NUV of 0.153 and 0.227 mag 
Following this, we obtained FUV and NUV flux densities of $2.1 \pm 0.1 \times 10^{-4}$ Jy and $10.3 \pm 0.3 \times 10^{-4}$ Jy, corresponding to luminosities of $1.2 \pm 0.1 \times 10^{7}\mbox{L}_\odot$ and $3.8 \pm 0.1 \times 10^{7}\mbox{L}_\odot$, respectively.   
% For comparison, if we applied the Galactic extinction for FUV and NUV same as \citet{Hao2011}, the difference for the flux densities would only vary about $ 1\%$.  
Next, we corrected the ultraviolet fluxes for the intrinsic dust extinction using
\begin{subequations}
\begin{align}
 & F({\mbox{FUV}})_{\mbox{corr}} = F({\mbox{FUV}})_{\mbox{obs}} + 0.46 F({\mbox{TIR}}), \label{eq:FUV_cor_TIR} \\
 & F({\mbox{FUV}})_{\mbox{corr}} = F({\mbox{FUV}})_{\mbox{obs}} + 3.89 \nu f_\nu(22\mu{\mbox{m}}) \label{eq:FUV_cor_22}
%  & \nu f_\nu({\mbox{FUV}})_{\mbox{cor}} = \nu f_\nu({\mbox{FUV}})_{\mbox{obs}} + 0.46 F({\mbox{TIR}}), \\
%  & \nu f_\nu({\mbox{FUV}})_{\mbox{cor}} = \nu f_\nu({\mbox{FUV}})_{\mbox{obs}} + 3.89 \nu f_\nu(22\mu{\mbox{m}})
\end{align}
\end{subequations}
and
\begin{subequations}
\begin{align}
 & F({\mbox{NUV}})_{\mbox{corr}} = F({\mbox{NUV}})_{\mbox{obs}} + 0.27 F_\nu ({\mbox{TIR}}), \label{eq:NUV_cor_TIR} \\
 & F({\mbox{NUV}})_{\mbox{corr}} = F({\mbox{NUV}})_{\mbox{obs}} + 2.26 \nu f_\nu(22\mu{\mbox{m}}) \label{eq:NUV_cor_22}
% & \nu f_\nu({\mbox{NUV}})_{\mbox{cor}} = \nu f_\nu({\mbox{NUV}})_{\mbox{obs}} + 0.27 F({\mbox{TIR}}), \\
% & \nu f_\nu({\mbox{NUV}})_{\mbox{cor}} = \nu f_\nu({\mbox{NUV}})_{\mbox{obs}} + 2.26 \nu f_\nu(22\mu{\mbox{m}})
\end{align}
\end{subequations}
and then converted the extinction-corrected ultraviolet fluxes to SFRs using 
%, and we obtained FUV fluxes of $3.85\pm 0.24\times 10^{14}$ Jy Hz corrected using TIR flux and $6.20\pm 0.37\times 10^{14}$ Jy Hz corrected using mid-infrared flux and NUV fluxes of $2.27\pm 0.14\times 10^{14}$ Jy Hz corrected using TIR flux and $3.61\pm 0.21\times 10^{14}$ Jy Hz corrected using mid-infrared flux.  Using
\begin{equation}
  \frac{\mbox{SFR}(\mbox{FUV})}{\mbox{M}_\odot\mbox{ yr}^{-1}}=5.34\times10^{-17}
  \left[\frac{D}{\mbox{ Mpc}}\right]^{2}
  \left[\frac{F(\mbox{FUV})}{\mbox{Jy Hz}}\right]
  % 4.47\times10^{-44} \frac{\nu L_\nu(\mbox{FUV})}{\mbox{erg}\mbox{ s}^{-1}}
\label{eq:sfr_FUV}
\end{equation}
and
\begin{equation}
  \frac{\mbox{SFR}(\mbox{NUV})}{\mbox{M}_\odot\mbox{ yr}^{-1}}=8.09\times10^{-17}
  \left[\frac{D}{\mbox{ Mpc}}\right]^{2}
  \left[\frac{F(\mbox{NUV})}{\mbox{Jy Hz}}\right]
  % 6.76\times10^{-44} \frac{\nu L_\nu(\mbox{NUV})}{\mbox{erg}\mbox{ s}^{-1}}
\label{eq:sfr_NUV}
\end{equation}
given by \citet{Hao2011}. Finally, we obtained the SFRs listed in Table~\ref{tab:SFR}.  
% SFRs$(\mbox{FUV})$ of $1.93 \pm 0.12$ and $3.04 \pm 0.17 {\mbox{M}_\odot\mbox{ yr}^{-1}}$ corrected using TIR and mid-infrared emission, and SFRs$(\mbox{NUV})$ of $1.72 \pm 0.10$ and $2.68 \pm 0.15 {\mbox{M}_\odot\mbox{ yr}^{-1}}$ corrected using TIR and mid-infrared emission, respectively. 

The combinations of ultraviolet continuum fluxes corrected using TIR flux yield SFRs about 60 per cent of the ALMA free-free SFR.  The primary sources of uncertainty in these calculations are the uncertainties in the extinction correction, which are $\sim 0.1$ dex \citep{Hao2011}.  Given this, the difference between the ALMA free-free SFR and the SFR form the combined ultraviolet and TIR fluxes is statistically significant.  As was the case for the SFRs based on the extinction-corrected $\ha$ flux, the SFRs from the extinction-corrected ultraviolet flux may yield inaccurate results in this situation because the central starburst in NGC~1808 is heavily dust-obscured.  The observed TIR to FUV luminosity ratio is 3.32 within the central region, which exceeds the maximum of the applicability of 2.39 give by \citet{Hao2011}. 

However, the SFRs from the ultraviolet continuum emission corrected using mid-infrared emission are broadly consistent with the ALMA free-free SFR.  This could be in part because the 22~$\mu$m band is more strongly associated with the dust that is absorbing and re-radiating energy from star-forming regions.  We also note that, in Equations \ref{eq:FUV_cor_TIR}-\ref{eq:NUV_cor_22}, the observed ultraviolet fluxes are only equivalent to less than 1 per cent of the infrared terms, so the conversions from the extinction-corrected ultraviolet fluxes to SFRs are effectively indirect conversions from infrared fluxes to SFRs.  In other words, the agreement between the ultraviolet SFRs corrected using mid-infrared emission and the ALMA free-free SFR in some way arises from the agreement between the infrared SFRs and the ALMA free-free SFR.  This overall situation seems somewhat awkward and highlights the difficulty in using Equations \ref{eq:FUV_cor_22} and \ref{eq:NUV_cor_22} for calculating SFRs for dusty starburst galaxies.  

The close match between the SFRs from the ultraviolet flux corrected using the 22~$\mu$m flux and the SFRs from the ALMA free-free emission could actually be the result of two separate biases cancelling out.  It may be the case that using infrared fluxes to correct ultraviolet fluxes will generally yield low SFRs for heavily obscured systems like the centre of NGC~1808.  However, as we noted in Section~\ref{subsubsec:SFR_monIR}, the dust in NGC~1808 appears to be relatively warm compared to typical galaxies, which means that the 22~$\mu$m flux density will yield a higher extinction correction than the TIR flux.  This could cancel out effects that bias the SFRs from extinction-corrected ultraviolet emission to lower values.

The discrepancy between SFRs from the ultraviolet fluxes corrected using TIR and corrected using mid-infrared fluxes is similar to that found from the $\ha$ results in Section~\ref{subsec:SFR_ha}, but the corrected ultraviolet fluxes yield systematically higher SFRs than the corrected $\ha$ fluxes.  The calibration coefficients of the infrared term used for correcting the ultraviolet fluxes (Equations \ref{eq:FUV_cor_TIR}-\ref{eq:NUV_cor_22}) is significantly higher than that used for correcting the $\ha$ fluxes (Equations \ref{eq:ha_cor_TIR}-\ref{eq:ha_cor_22}), which makes the SFRs based on extinction-corrected ultraviolet emission more dependent on the infrared emission.  
% The difference between SFRs from ultraviolet and $\ha$ fluxes may arise from different efficiencies of these calibration methods when applied to the dust-obscured situation that exceeds their applicability ranges.  

NUV continuum emission is expected to be less reliable than FUV continuum emission as an SFR tracer because stars older than 100 Myr may produce a significant fraction of the total NUV emission from a typical galaxy \citep[e.g.,][]{Hao2011}.  However, the SFRs that we derived from the extinction-corrected NUV and FUV fluxes are similar.  This is most likely because, as we discussed above, the FUV and NUV terms in Equations \ref{eq:FUV_cor_TIR}-\ref{eq:NUV_cor_22} are negligible compared to the corresponding infrared terms.
% The young stellar populations with ages between 10 and 100 Myr can also contribute ultraviolet continuum emission aside from the photoionzing stars with ages less than 10 Myr \citep{Kennicutt2012}. However, we cannot judge the based on the comparisons of SFRs from ultraviolet fluxes and $\ha$ fluxes because of the negelible fraction of ultraviolet fluxes and $\ha$ fluxes compared to infrared terms in above conversions.   
% On the other hand, the different stellar ages traced by ultraviolet continuum (0-100 Myr) and $\ha$ line emission (0-10 Myr) \citep{Kennicutt2012}, and the other young stellar populations (e.g., 10-100 Myr) could contribute additional ultraviolet continuum emission. 

%\subsection{Comparisons of ALMA and other SFR measurements}
\section{Conclusions} \label{sec:conclusion}

We have presented ALMA observations of 85.69 and 99.02 GHz continuum emission and H42$\alpha$ and H40$\alpha$ line emission at their rest frequencies from NGC~1808.  The millimetre continuum emission primarily originates from a nucleus and a fainter star-forming region to the south of the nucleus.  We detected H40$\alpha$ and H42$\alpha$ lines from the southern star-forming region but only detected H42$\alpha$ line from the nucleus, which might be explained by the higher noise levels in the H40$\alpha$ spectra and the relatively low flux of the recombination lines generally.  It is also possible that the H42$\alpha$ line emission from the galactic nucleus is blended with the $c$-C$_3$H$_2$ line at 85.65~GHz. 

Through an analysis of the SED from 1 to 360~GHz, we demonstrate that free-free emission is the dominant source of emission at 85-100~GHz, with synchrotron emission mainly found at lower frequencies and thermal dust emission as the preeminent source at higher frequencies.  Given this and given the lower reliability of the results from the hydrogen recombination lines, we primarily rely on the free-free continuum to calculate SFRs.

However, we can still reliably calculate electron temperatures within NGC~1808 based on the ratios of the recombination lines to the continuum emission.  The average electron temperature for the central 22~arcsec (1~kpc) diameter region, which encompasses the nuclear and southern millimetre sources, is $5200\pm 500$~K.  This is comparable to that of the centre of the Milky Way and NGC~4945 and the northern nucleus of NGC~3256.

The SFR within the central 22~arcsec derived from the 85.69 and 99.02~GHz free-free emission is $3.1\pm0.3$~M$_\odot$~yr$^{-1}$.  We treat this as the best possible measure of the SFR within this central region and compare it to other star formation metrics to test their effectiveness and, if they produce differing results, to understand how other physical processes could affect the accuracy of these metrics.

The SFR from the ALMA data is statistically consistent with what we measure using the infrared emission, probably because the region is heavily dust obscured, so most of the bolometric energy from the region emitted as infrared light.  However, because the dust within NGC~1808 is relatively hot compared to the typical star-forming and starburst galaxies, the SFRs from individual infrared bands decrease as wavelength increases, which indicates the limited efficacy of these bands by themselves to measure SFRs.  The SFR from ALMA also falls in the range of SFRs measured from previous Br$\gamma$ line emission by \citet{Krabbe1994}, \citet{Kotilainen1996} and \citet{Busch2017}, although the first two references provide a broad range of SFRs (as a result of considering a broad range of IMFs).  Still, this may demonstrate that dust extinction in the near-infrared bands can still be straightforwardly corrected to derive accurate SFRs, which may not necessarily be the case with more obscured nuclear starbursts like the one in NGC~4945 \citep{Bendo2016}.

The SFRs from the 1.5~GHz emission, PAH emission and extinction-corrected $\ha$ are significantly lower than the SFR from the ALMA free-free emission.  The 1.5~GHz SFR is probably low because synchrotron radiation produced by cosmic rays should generally appear broader than the star-forming regions themselves \citep{Murphy2006}.  The PAH SFR is low most likely because PAH emission is generally expected to be suppressed relative to other star formation tracers in strong starbursts \citep{Helou2004,Calzetti2005,Calzetti2007,Bendo2006,Bendo2008,Prescott2007}.  The extinction-corrected $\ha$ SFR is low possibly because the central region is heavily dust obscured and because the equations that are commonly used are not designed for such extremely obscured regions \citep{Kennicutt2009}.

The combined ultraviolet and 22~$\mu$m metrics yield accurate SFRs.  However, these SFRs are effectively based on just the 22~$\mu$m emission; in the equations we used to calculate SFRs, the ultraviolet terms were negligible compared to the 22~$\mu$m term.  The ultraviolet emission corrected with the TIR fluxes yielded lower SFRs.  Even though the extinction corrections using 22~$\mu$m and TIR fluxes were both derived by \citet{Hao2011}, the corrections with the TIR fluxes probably yield lower SFRs because the equations are based on galaxies with colder dust temperatures than what is seen in NGC~1808.

% Regarding the ambiguous presence of the AGN in the nucleus of NGC~1808,
% Some studies have attempted to determine whether an AGN is present in NGC~1808 \citep[e.g.,][]{Awaki1996,Jimenez-Bail2005,Audibert2021}.  However, the absence of any broad line emission, the dominance of the free-free emission at millimetre continuum emission and the relatively low radio SFR in the centre of NGC~1808 all suggest that there is no clear evidence for a strong AGN in this galaxy.

Whether an AGN is present in NGC~1808 is still under debate \citep[e.g.,][]{Dopita2015,Busch2017,Audibert2021}.  In our SED analysis, we found that free-free emission was the dominant source of emission in ALMA Band 3, whereas if an AGN was present, we may expect additional synchrotron emission.  Additionally, we obtained a relatively low SFR from the 1.5~GHz band, and the SFRs from the infrared data were similar to what was derived from the ALMA free-free continuum emission.  If an AGN was present (or if it was a significant energy source), we would expect the AGN to produce additional flux in these bands, which would have yielded higher SFRs than what we obtained from the free-free emission.  It is still uncertain whether the enhancement of synchrotron radiation of the galactic nucleus compared to the southern star-forming region as seen in our SED analysis is indicative of the presence of an AGN or merely a normal variation in synchrotron emission levels within the galaxy.  
Unfortunately, we have very few other analyses of the radio to far-infrared SEDs of nearby galaxies that we can use in comparison to these results to show whether the synchrotron emission is truly enhanced within the nucleus of this galaxy.  However, we will be able to investigate this topic further with the upcoming Square Kilometre Array (SKA) and Next Generation Very Large Array (ngVLA).
% the absence of any broad millimetre line emission, 
%% Wang 2023 no broad RRL (H35alpha & H36alpha) in type II AGN, Circinus galaxy
%%% it is hopeless to try to detect broad RRL emission in local AGNs with current facilities. 
%% Izumi 2016 no broad or narrow H26alpha line in type II AGN, NGC1068

This study further demonstrates the effectiveness of using ALMA observations of free-free and hydrogen recombination line emission to study star formation in the centre of starburst galaxies and examine the applicability of other star formation metrics.  In particular, millimetre recombination lines from other galaxies are rarely detected using ALMA, so any new detection is notable in and of itself.  Studies based on both new ALMA observations and existing archival data should be used to expand the number of galaxies with known line detections as well as to measure the free-free emission in more galaxies so as to better understand the nature of star formation in the extragalactic universe.

%The images of multi-wavebands reveal that the morphologies of this galaxy differ from different wavebands.  The millimetre continuum emission originates from a nucleus and a southern region, which are also visible in the 1.51 GHz continuum emission.  In contrast, the $\ha$ line emission traces a more extended disc, and the ultraviolet continuum emission traces a few unresolved bright sources with an arc structure.  Their morphologies are different from those of the millimetre continuum emission because they are heavily affected by dust obscuration in the centre of this galaxy.

% The SFRs from the combinations of $\ha$ line and infrared emission are significantly lower than what we measured using the ALMA data because of the heavy dust obscuration in the central starburst. 

%The negative dependence of the SFRs from monochromatic infrared emission on wavelength suggests that the dust within NGC~1808 is relatively hot compared to the typical star-forming and starburst galaxies.  Such a hot dust in the central starburst may also cause systematic discrepancies between the SFRs from the extinction-corrected $\ha$ line flux (or ultraviolet fluxes) using the total infrared flux and using the mid-infrared flux.

\section*{Acknowledgements}

%The Acknowledgements section is not numbered. Here you can thank helpful colleagues, acknowledge funding agencies, telescopes and facilities used etc. %Try to keep it short.

% This work is supported by the Strategic Priority Research Program of Chinese Academy of Sciences (Grant No. XDB 41000000), the National Science Foundation of China (NSFC, Grant No. 12233008, 11973038), the China Manned Space Project (No. CMS-CSST-2021-A07), the Cyrus Chun Ying Tang Foundations, the Frontier Scientific Research Program of Deep Space Exploration Laboratory, and the 111 Project for "Observational and Theoretical Research on Dark Matter and Dark Energy" (B23042).  GC also acknowledges support from the SKA exchange programme of STFC and CSC (No. 201906340241).  GJB is supported by STFC grant ST/T001488/1.  GAF acknowledges support from the Collaborative Research Centre 956, funded by the Deutsche Forschungsgemeinschaft (DFG) project ID 184018867.  

GC acknowledges support from the SKA exchange programme of STFC and CSC (No. 201906340241).  % the China Scholarship Council 
GJB is supported by STFC grant ST/T001488/1.
GAF acknowledges support from the Collaborative Research Centre 956, funded by the Deutsche Forschungsgemeinschaft (DFG) project ID 184018867.  
XK acknowledges support from the Strategic Priority Research Program of Chinese Academy of Sciences (Grant No. XDB 41000000), the National Science Foundation of China (NSFC, Grant No. 12233008, 11973038), the China Manned Space Project (No. CMS-CSST-2021-A07), the Cyrus Chun Ying Tang Foundations, the Frontier Scientific Research Program of Deep Space Exploration Laboratory, and the 111 Project for "Observational and Theoretical Research on Dark Matter and Dark Energy" (B23042).
This paper makes use of the following ALMA data: ADS/JAO.ALMA\#2016.1.00562.S, 2012.1.01004.S, 2013.1.00911.S, and 2017.1.00984.S. ALMA is a partnership of ESO (representing its member states), NSF (USA) and NINS (Japan), together with NRC (Canada), MOST and ASIAA (Taiwan), and KASI (Republic of Korea), in cooperation with the Republic of Chile.  The Joint ALMA Observatory is operated by ESO, AUI/NRAO and NAOJ.

%%%%%%%%%%%%%%%%%%%%%%%%%%%%%%%%%%%%%%%%%%%%%%%%%%
\section*{Data Availability}

%The inclusion of a Data Availability Statement is a requirement for articles published in MNRAS. Data Availability Statements provide a standardised format for readers to understand the availability of data underlying the research results described in the article. The statement may refer to original data generated in the course of the study or to third-party data analysed in the article. The statement should describe and provide means of access, where possible, by linking to the data or providing the required accession numbers for the relevant databases or DOIs.

The reduced, calibrated and science-ready ALMA data is available from the ALMA Science Archive at \url{https://almascience.eso.org/alma-data/archive} .



\bibliographystyle{mnras}
\bibliography{NGC1808} % if your bibtex file is called example.bib

%\begin{thebibliography}{99}
%\end{thebibliography}

%%%%%%%%%%%%%%%%%%%%%%%%%%%%%%%%%%%%%%%%%%%%%%%%%%

%%%%%%%%%%%%%%%%% APPENDICES %%%%%%%%%%%%%%%%%%%%%

% \appendix
% \section{Some extra material}
%If you want to present additional material which would interrupt the flow of the main paper, it can be placed in an Appendix which appears after the list of references.

%%%%%%%%%%%%%%%%%%%%%%%%%%%%%%%%%%%%%%%%%%%%%%%%%%

% Don't change these lines
\bsp	% typesetting comment
\label{lastpage}
\end{document}